\documentclass[twoside,12pt]{article}
\usepackage{epsfig}
\usepackage{amssymb}
\newcommand{\be}{\begin{equation}}
\newcommand{\ee}{\end{equation}}
\newcommand{\bea}{\begin{eqnarray}}
\newcommand{\eea}{\end{eqnarray}}
\newcommand{\nn}{\nonumber}
\newcommand{\lb}{\label}

\begin{document}
\markright{\LaTeX\ guidelines for Elsevier Major Reference Works}

\parindent 0mm
\parskip 6pt

% Start with the article header information
% Include all author details, including postal and e-mail addresses

\title{Continuous monitoring and the introduction of a classical level in Quantum Theory}

\author{G. M. Prosperi\\
Dipartimento di Fisica dell'Universit\'a di Milano\\
I. N. F. N. , sezione di Milano
Via Celoria 16, I20133 Milano, (Italy)}

\date{Aug 2015}

\maketitle

\begin{abstract}
In ordinary Quantum Mechanics only ideally instantaneous observations of a quantity 
or a set of compatible quantities are usually considered. In an old paper of our group in 
Milano a formalism was introduced for the continuous monitoring of a system during a 
certain interval of time in the framework of a somewhat generalized approach to Q. M. 
The outcome was a distribution of probability on the space of all the possible continuous 
histories of a set of quantities to be considered as a kind of coarse grained 
approximation to some ordinary quantum observables commuting or not.
The main aim was the introduction of a classical level in the context of 
Quantum Mechanics, treating formally a set of basic quantities {\it to be considered as
beables} in the sense of Bell as continuously taken under observation. However the effect 
of such assumption was a permanent modification of the Liouville-von Neumann equation 
for the statistical operator by the introduction of a dissipative term which is in conflict 
with basic conservation rules in all reasonable models we had considered. Difficulties 
were even encountered for a relativistic extension of the formalism.
In this paper I propose a modified version of the original formalism which seems to overcome 
both difficulties. First I study the simple models of an harmonic oscillator and 
a free scalar field in which a coarse grain position and a coarse grained field 
respectively are treated as beables. Then I consider the more realistic case of Spinor 
Electrodynamics in which only certain coarse grained electric and magnetic fields and no matter 
related quantities are introduced as classical variables.

\end{abstract}

%11111111111111111111111111111111111111111111111111111111111

\section{Introduction}
\quad In reference \cite{BLP} a general formalism was introduced for the treatment 
of the {\it continuous monitoring} of a quantity or a set of quantities in the framework of 
Quantum Mechanics. The formalism turns out to be strictly related to a more particular 
one previously proposed by E. B. Davies for the observation of the counting times on a 
system of counters \cite{Davis}. This is in the context of the generalized formulation of Quantum 
Mechanics (GQM) based on the concept of positive operator valued measures (p.o.m.) and 
operation valued measures or instruments, originally proposed by G. Ludwig, E. B. Davies,
S. Holevo \cite{Davis}-\cite{Holevo}. The outcome is a distribution of probability 
on the set of all the possible continuous histories  of the monitored quantities in the 
considered time interval. The class of event subsets is characterized in terms of time 
average of the monitored quantities of the type
\be
\lb{1.0}
a_h^s(t) =\int dt' h(t-t')\,a^s(t')\,,
\ee
the $h(t)$s being appropriate weight functions (generalized stochastic process).

\quad Later various alternative formulations of the theory have been given and various 
aspects developed with interesting applications, particularly in the field of Quantum 
Information and Optics (for a recent presentation see e. g.  \cite{Barchielli} and 
references therein).

\quad The original purpose of \cite{BLP} was, however, to obtain a modification of 
ordinary Quantum Theory, in which an intrinsic classical level for some basic macroscopic 
quantities could be introduced, to solve consistently the problem of interpretation of the 
Theory in the sense of Bohr and of Von Neumann. The idea was that certain basic 
quantities should be chosen once for ever as an additional postulate and they should be 
thought as having at any time a well defined value, considered {\it beables} in the sense of Bell, by 
treating them formally as continuously observed. Then any other observation on a 
microscopic subsystem should be expressed in terms of the modifications that its 
interaction with the remaining part of the world  produces in the value of such 
basic macroscopic quantities. Obviously the theory  remain statistic and to any 
possible evolution of the basic quantities a precise probability is assigned. Note that, even 
if in a completely different mathematical framework, similar conceptual ideas seem to be at the 
basis of the so called theory of the {\it consistent histories} \cite{dechyst}. With reference to 
them see however also ref. \cite{despagnat}.  

\quad A significant property of the modified theory is that integrating  over all 
possible histories of the basic quantities in a given time interval $(t_0,  t_F)$  is
equivalent  to introduce a dissipative term in the Liouville-Von Neumann evolution 
equation for the statistical operator in  the Schroedinger picture. A particular interesting 
case of such a modified Liouville-Von Neumann equation is
\be  
{\partial \hat \rho _S(t) \over \partial t}= -i [\hat \rho_S(t), \hat H]
                -\sum_j \alpha_j [\hat A^j,[\hat A^j,\hat \rho_S (t)]],
\label{1.1}
\ee
where $\hat A^1,\, \hat A^2,\, \dots\, $ denote hermitian operators not necessarily 
each other commuting and $\alpha_1, \alpha_2, \dots $  positive constants. Alternatively in Heisenberg 
picture, which we find more convenient in this paper, this corresponds to ascribe a time dependence 
to the statistical operator according to the equation
\be  
{\partial \hat \rho(t) \over \partial t}= 
                -\sum_j \alpha_j [\hat A^j (t),[\hat A^j (t),\hat \rho (t)]].
\label{1.2}
\ee
\quad In the case the monitored quantities $a^1,\,a^2,\,\dots$ may be considered 
simultaneous coarse grained approximations of the ordinary quantum observables 
$A^1,\, A^2,\, \dots\, $ with  $\langle a^s(t)\rangle=\langle \hat A^s(t) \rangle$ and 
the variance $\langle(a_h^s(t)-  \langle a_h^s(t) \rangle)^2\rangle $ expressed as the 
sum of an intrinsic term independent of $\hat \rho$ and a minor modification of the 
ordinary quantum variance for $A^s (t)$. To $a^1,\,a^2,\,\dots$ in the following we 
shall conventionally refer as the  macroscopic $A^1,\, A^2,\, \dots\, $ 

\quad  Notice that eqs. (\ref{1.1}) or (\ref{1.2}) are trace and positivity preserving. 
The dissipative term expresses the permanent effect of the modification introduced in the 
theory (formally the perturbation produced by the continuous monitoring), even when any 
information on the mentioned basic quantities is completely disregarded. The above 
equations make also our theory in contact with theories that introduce 
ad hoc dissipative terms, as a noise, to simulate the  interaction of the apparatus with 
an environment (see in particular in this connection ref. \cite{Zeh}) or theories that 
want introduce an intrinsic progressive decoherence and a spontaneous collapse at a more 
fundamental level. Theories of the latter type received considerable attention in the last 
thirty years (see e. g. \cite{collapse} for general revues and complete references; a 
small representative sample, corresponding to various point of view, is reported in refs.
\cite{GRW}-\cite{diosi}).

\quad  In our context specific examples of choice of the basic macroscopic quantities in simple models 
may be the macroscopic density of particle in a non-relativistic second quantization 
theory with self-interaction,  the particle distribution functions in the classical 
phase space for a system of electrons and protons \cite{Prosperi}, a macroscopic 
field for a self-interacting quantum scalar field \cite{BLP}. Unfortunately in all 
examples we have considered eqs. (\ref{1.1}) or (\ref{1.2}) turn out to be 
in conflict with important conservation rules (the energy conservation rule in 
particular) at a level that does not seem compatible with the matter stability as 
presently established \cite{BLP},  \cite{Prosperi} (in this connection see however even 
\cite{GRW}). There are also difficulties to a covariant extension of the formalism to a 
vector field, like the electro-magnetic field, or to a tensor one and, even in the case 
of the scalar field, the additional term in (\ref{1.1}, \ref{1.2}) would be ill defined.

\quad  All difficulties seem to be related to the requirement that $\alpha_1,\alpha_2,\dots$ 
were  positive. In this paper we want to show that it is possible to release such requirement and  
significant models can be constructed in which by an appropriate choice of the 
operators $\hat A^1, \hat A ^2, \dots $, of the constants $\alpha_1, \alpha_2, 
\dots $ and of the weight functions $h(t)$ a consistent probability distribution for the 
histories of related basic quantities can be defined and the conservation rules 
respected. We have to assume that $\hat\rho$ is trace 1 and positive at some initial time 
$ t_0 $ but  it is not necessary that the positivity of $\hat \rho (t)$ is preserved 
in time to have a positive probability for the quantities of interest.

\quad Among the above models, it is of particular interest the case of spinor QED in 
which the basic quantities to be considered as {\it classical} or {\it beables} are the macroscopic 
components of the electromagnetic field, but nothing concerning matter. In this case eq. (\ref{1.2}) 
becomes
\be
\label{1.3}
{\delta \hat \rho [\sigma] \over \delta \sigma (x)}= {\gamma \over 16}\, 
\left [\hat F_{\mu \nu}(x),\,\left [\hat F^{\mu
      \nu}(x),\hat\rho [\sigma]\,\right ] \right ]\,,
\ee  
where $\sigma$ is a space-like surface passing across $x$. Note that the form of this 
equation is practically completely determined by Lorentz and gauge invariance requirements and 
that it corresponds to $\alpha_1,\,\alpha_2,\, \dots $ not all positive. Furthermore
as in any field theory the averages (\ref{1.0}) in terms of which the probabilities
are defined have to be replaced by expressions of the type
\be
\lb{1.4}
f_h^{\mu\nu}(x)=\int d^4x\,h(x-x')\, f^{\mu\nu}(x')
\ee
and the restriction on the $h(x)$s consists in the requirement that  only time like wave 
vectors $k$ occur in the Fourier transform $\tilde h(k)$ or that they are dominant. 
 
\quad The plan of the paper is the following one. In section 2 we shall review 
the formalism of the continuous monitoring, manly to establish notations, and recall the 
important notion of functional generator. In sect 3 we shall consider the case of 
the harmonic oscillator in the original formulation, assuming as basic quantity the 
macroscopic position and show the positivity of the corresponding probability by using 
path integral techniques; then we extend discussion to the case of a free scalar field. 
In sect. 4 we discuss the problem of the conservation rules and in sect. 5 we show how 
the formalism can be consistently modified in order to dispose of the corresponding 
violation always for the harmonic oscillator and the scalar field. In sect. 6 we 
consider the mentioned more significant case of spinorial QED. Finally in sect. 7 we 
summarized the results and try to make some  conclusions and additional remarks.
Some technicalities are confined in the appendices.

%2222222222222222222222222222222222222222222222222222222222

\section{Continuous monitoring of a set of quantities}

\quad   In GQM a set of \textit{compatible observables} 
$A \equiv (A^1, A^2, \dots A^p)$ is associated to a normalized {\it effect} or 
\textit{positive operator valued measure} (p.o.m.) $\hat F_A(T)$ and the apparatus
$S_A$ for observing them to an \textit{instrument} or \textit{operation valued measure} 
(o.v.m.) $\mathcal F_{S_A}(T)$, $T$ being a Borel subset of the real space $\Re^p$ of 
all possible values of $A$.

\quad  That is, $\hat F_A(T)$ and $\mathcal F_{S_A}(T)$ are a positive operator on the  
Hilbert space $\mathcal{H}$ associated to the system and a mapping of the set 
of the trace class operators in itself, respectively, satisfying the relations 
\begin{equation}
\label{2.1}
 \hat F(\cup _{j=1}^n T_j)=\sum _{j=1 ^n} \hat F(T_j) \qquad\ 
 {\rm and} \qquad 
\mathcal F(\cup _{j=1}^n T_j)=\sum _{j=1 ^n} \mathcal F(T_j)   \,,
\end{equation}
if
\begin{equation}
\label{2.2}
T_i\cap T_j = 0
\end{equation}
and
\begin{equation}
\label{2.3}
 \hat F_A(\Re ^p) = \hat I \qquad {\rm Tr}[\mathcal F_{S_A}(\Re^p) \hat X]=
{\rm Tr}\hat X \,.
\end{equation}
 Further they must be related each other by the equation
\begin{equation}
\label{2.4}
\hat F_A(T)=\mathcal F_{S_A}^\prime(T) \hat I \,,
\end{equation}
where by $\mathcal F ^\prime $ we denote the dual mapping of $\mathcal F$, defined by
the equation
\be
\lb{2.4a}
{\rm Tr}\left[\hat B \mathcal F  \hat X \right]=
          {\rm Tr}\left[ \left (\mathcal F^\prime \hat B \right) \hat X \right]\,,
\ee
$\hat X$ being an arbitrary trace class operator and $\hat B$ an arbitrary bounded 
operator.
So
\begin{equation}
\label{2.5}
 {\rm Tr}[\hat F_{A}(T) \hat X]= 
{\rm Tr}[\mathcal F_{S_A}(T)\hat X] \,.
\end{equation}

\quad As we told, we shall find convenient to work in Heisenberg picture. Then we have
\begin{eqnarray}
&&\qquad \quad \hat F_A(T,t)=e^{iHt}\hat F_{A}(T) e^{-iHt}\\ \nonumber
&& \mathcal F_{S_A}(T,t)\hat X = e^{iHt}[\mathcal F_{S_A}(T)(e^{-iHt}\hat
  Xe^{iHt})] e^{-iHt}
\label{2.6} 
\end{eqnarray}
and the probability of observing $A \in T$ at the time $t$ is
\begin{equation}
\label{2.7}
P(A \in T,t| \rho)={\rm Tr}[\hat F_A(T,t)\hat \rho]
={\rm Tr}[\mathcal F_{S_A}(T,t)\hat \rho]\,,
\end{equation}
where $\hat\rho$ denotes the statistical operator representing the state of the system 
(a priori a mixture state), which we usually indicate by $\rho$.

\quad The reduction of the state as consequence of having observed $A \in T$ at 
the time $t_0$ by the apparatus $S_A$ must be written as
\begin{equation}
\label{2.8}
\hat \rho \to \mathcal F_{S_A}(T,t_0)\hat \rho /
{\rm Tr}[\mathcal F_{S_A}(T,t_0)\hat \rho]\,.
\end{equation}
\quad Notice
\begin{equation}
\label{2.9}
\langle A^j \rangle = {\rm Tr}[\hat A^j(t) \hat \rho ]
\end{equation}
with 
\begin{equation}
\label{2.10}
\hat A^j(t) = e^{iHt} \hat A^j  e^{-iHt} \qquad {\rm and} \qquad
\hat A^j = \int _{\Re ^p} d \hat F(a) a^j \,.
\end{equation}
The operators $\hat A^j$ are Hermitian but generally they do not commute. Such a set of 
generalized compatible observables can be interpreted as corresponding to an approximate
simultaneous measurement of possibly incompatible ordinary observables 
$\hat A_1,\,\hat A_2,\,\dots$ 

\quad Now let us assume that we make repeated {\it independent}  observations on $A$ at 
subsequent times $t_0, t_1, \dots t_N$. Combining eqs. (\ref{2.7}) and (\ref{2.8}) the 
\textit{Joint probability} of observing a \textit{a sequence of results} for $A$  can be 
written 
\begin{eqnarray}
&& P(A \in T_N, t_N; \dots A \in T_1, t_1; A \in T_0, t_0|\rho) = \qquad  \nn \\
&& \quad ={\rm Tr}[\mathcal F_{S_A}(T_N,t_N)\dots \mathcal F_{S_A}(T_1,t_1)
\mathcal F_{S_A}(T_0,t_0) \hat \rho]
\label{2.11}
\end{eqnarray}  
Notice that
\begin{equation}
\mathcal F( T_N, t_N; \dots ; T_1, t_1; T_0, t_0) =
\mathcal F_{S_A}(T_N,t_N)\dots \mathcal F_{S_A}(T_1,t_1)
\mathcal F_{S_A}(T_0,t_0))
\label{2.12}
\end{equation}  
and
\begin{equation}
\hat F( T_N, t_N; \dots ;  T_1, t_1; T_0, t_0) =
\mathcal F_{S_A}^\prime (T_0,t_0) \mathcal F_{S_A}^\prime (T_1,t_1)
\dots \mathcal F_{S_A}^\prime (T_N,t_N)) \hat I
\label{2.13}
\end{equation}  
define an instrument and a p.o.m. on  a real space with $p(N+1)$ dimensions $\Re^{p(N+1)}$\,.

\quad Then
\begin{eqnarray}
&& P(A \in T_N, t_N; \dots ; A \in T_0, t_0|\rho)=  \nonumber \\
&&  \qquad \qquad = {\rm Tr}[\mathcal F( T_N, t_N; \dots ; T_0, t_0)
\hat \rho]  \nn \\
&& \qquad \qquad = {\rm Tr}[\hat F( T_N, t_N; \dots ; T_0, t_0) \hat \rho]\,.
\label{2.14}
\end{eqnarray}  
So in GQM the observation of a sequence of results at certain successive times can
be put on the same foot as the observation of $A$ at a single time.

% 2.1 2.1 2.1 2.1 2.1 2.1 2.1 2.1 2.1 2.1 2.1 2.1 2.1 2.1 2.1 2.1 2.1 2.1 

\subsection{Continuous monitoring}

\quad On analogy with above let us consider the limit case of a set of quantities
continuously kept under observation.
\footnote{Note that in the framework of 
ordinary quantum mechanics, in which only exact observations are considered 
corresponding to projection valued measures, there is a negative theorem in 
connection with this problem, usually recalled as Zeno's theorem. According 
to such theorem, if we make repeated observations of the same quantity and let 
go to 0 the interval of time $\tau$ between two subsequent observations, the value 
of the quantity is frozen to its initial value and does not longer change with time. 
However, in GQM is possible to consider a double limit in which, as $\tau \to 0$, the 
observation in itself is made progressively less precise in such way that a 
finite result is attained. The discussion in this section can be thought in 
this perspective. The result is a generalized stochastic process in the 
sense of Gelfand, in which a set  of histories of finite measure of the quantities of 
interest has to be specified in  terms successive time averages of the type (\ref{1.0}) 
of the quantities of interest rather then of values assumed at definite times (see subsec. 2.3).}

\quad Let $\mathcal Y $ be the functional space of all possible \textit{histories} 
$\,\,a(t) \equiv  (a^1(t), a^2(t),\\  \dots \,a^p(t) )$ of 
a set of  quantities
in a reference time interval $(t_0,t_F)$ (where $t_F$ may be possible taken 
to $+\infty$) and $\Sigma$ the class of the measurable subsets of 
$\mathcal Y$ according to some definition to be specified later. Furthermore let us 
denote by $\Sigma_{t_a}^{t_b}\subset \Sigma$ the class of the measurable subsets of
$\mathcal Y$ corresponding to restrictions on the histories only in
the interval $(t_a,t_b)\subset (t_0,t_F) $.

\quad Then we assume that an \textit{instrument} and a related \textit{p.o.m.} 
\begin{equation}
\mathcal F(t_b,t_a; M) \qquad {\rm and}\qquad 
\hat  F(t_b,t_a; M)= \mathcal F^\prime(t_b,t_a; M) \hat I
\lb{2.1.1}
\end{equation}  
are defined, with $M\in \Sigma_{t_a}^{t_b}$ and we interpret
\begin{equation}
P(t_b,t_a; M) = {\rm Tr}[\hat F(t_b,t_a; M)\hat\rho (t_a)]=
{\rm Tr}[\mathcal F(t_b,t_a; M)\hat\rho(t_a)]
\label{2.1.2}
\end{equation}  
as the \textit{probability of observing} $a(t) \in M$
during the interval of time $(t_a,t_b)$ $\hat\rho(t_a)$ being the statistical operator 
at the time $t_a$ (see later). We shall call $ \hat F(t_b,t_a; M)$
and $\mathcal F(t_b,t_a; M)$ a positive operator valued stochastic process and an operation 
valued stochastic process (OVSP), respectively.

\quad  We assume that  $\mathcal F(t_b,t_a; M)$ satisfies the relation
\begin{equation}
\mathcal F(t_c,t_b; N)\mathcal F(t_b,t_a; M)=
\mathcal F(t_c,t_a; N\cap M) 
\label{2.1.3}
\end{equation}  
which as above it expresses the independence of the observation of $a(t)$ in successive 
intervals of time. Notice that
\begin{equation}
M\in \Sigma_{t_a}^{t_b}\,,\quad N\in \Sigma_{t_b}^{t_c}M
\quad \Rightarrow \quad N\cap M \in \Sigma_{t_a}^{t_c}\,.
\label{2.1.4}
\end{equation}  

\quad Under the above assumptions,
\begin{equation}
P(t_c,\,t_a;\, N\cap M) = {\rm Tr}[\hat F(t_c,t_b; N)\mathcal F(t_b,t_a; M)
\hat \rho (t_b)]
\label{2.1.5}
\end{equation}  
is the \textit{joint probability} of observing $a(t)\in M$ in the time
interval $(t_a,t_b)$ and $a(t)\in N$ in the interval
$(t_b,t_c)$. Then let us introduce the mapping
\begin{equation}
\mathcal G (t_b,t_a) \equiv \mathcal F(t_b,t_a; \mathcal Y) 
\label{2.1.6}
\end{equation}  
and notice that for the normalization requirement it must be trace preserving.
 
\quad By setting $M=\mathcal{Y}$ in (\ref{2.1.5}) we find
\begin{equation}
P(t_c,\,t_b;\,N)= P(t_C,\,T_A;\,N \cup \mathcal Y) = {\rm Tr}[\hat F(t_c,t_b; N)
\mathcal G(t_b,t_a) \hat \rho].
\label{2.1.7}
\end{equation}  
Then for comparison with (\ref{2.1.2}) we can set $\hat \rho(t_b)=\mathcal G(t_b,t_a) 
\hat \rho (t_a)$ and so the \textit{perturbation} produced on the system by its continuous 
observation introduces a kind of time dependence on the statistical operator in Heisenberg 
picture that is described by the action of the operator $\mathcal G(t_b,t_a)$.

\quad As we mentioned in the introduction we shall assume that our basic classical quantities 
are formally treated as continuously observed. In this perspective the OVSP 
$\mathcal F(t_b,t_a; M)$ must be chosen once for all and considered a
part of the theory and the action of the  related $\mathcal G(t_b,t_a)$  
 expresses the modification introduced in this way in the dynamics of the ordinary theory.

% 2.2 2.2 2.2 2.2 2.2 2.2 2.2 2.2 2.2 2.2 2.2 2.2 2.2 2.2 2.2 2.2  

\subsection{Characteristic functional operator}

\quad On analogy with the usual probability theory, we define 
the functional Fourier transform or {\it characteristic functional operator} 
(CFO) 
\begin{equation}
\mathcal G(t_b,t_a;[\xi(t)])=\int \mathcal {F}(t_b,t_a;\mathcal D_{\rm c }M) 
\exp \left \{- i \int_{t_a}^{t_b} dt \xi^s (t) a^s(t)\right \} \,,
\label{2.2.1}
\end{equation}  
where $\mathcal D_{\rm c} M $ denotes the measure of an elementary set in the functional space 
$\mathcal Y$ (the index ``c'' refers to the interpretation of $a(t)$ as a classical history, 
to distinguish the classical functional measure from the quantum path integral measure which 
shall be used in the following). The concept of CFO turns out to be very useful not only to 
study the properties of a given OVSP but even to construct such a structure. 

\quad Notice that in terms of $\mathcal G(t_b,t_a;[\xi(t)])$ assumption (\ref{2.1.3}) becomes
\begin{equation}
\mathcal G(t_c,t_b; [\xi(t)])\,\mathcal G(t_b,t_a; [\xi(t)])=
\mathcal G(t_c,t_a; [\xi(t)]) \,.
\label{2.2.2}
\end{equation}
Then, if we set
\begin{equation}
\mathcal G(t+dt,t; [\xi(t)])=1+ \mathcal K(t; \xi(t)) dt \,,
\label{2.2.3}
\end{equation}
we can write the differential equation
\begin{equation}
{\partial \over \partial t}\,\mathcal G(t,t_a; [\xi])= \mathcal K(t; \xi(t))\,
\mathcal G(t,t_a; [\xi]) \,,
\label{2.2.4}
\end{equation}
that we can formally solve as
\begin{equation}
\mathcal G(t_b,t_a; [\xi])={\rm T}\exp \int_{t_a}^{t_b} dt \mathcal K(t; \xi(t))\,,
\label{2.2.5}
\end{equation}
T being the usual time ordering prescription.

\quad Now let us observe that 
\be
\mathcal G(t_b,t_a; [0])=\mathcal {F}(t_b,t_a;\mathcal {Y}) \equiv \mathcal G(t_b,t_a).
\lb{2.2.6}
\ee
Then setting $\xi(t)=0$ in (\ref{2.2.4}) we obtain also 
\begin{equation}
{\partial \over \partial t}\,\mathcal G(t,t_a)= \mathcal L (t)\,
\mathcal G(t,t_a) 
\label{2.2.7}
\end{equation}
and
\begin{equation}
\mathcal G(t_b,t_a)={\rm T}\exp  \int_{t_a}^{t_b} dt \mathcal L (t)\,,
\label{2.2.8}
\end{equation}
with
\be
\mathcal L(t) = \mathcal K(t;0)\,.
\lb{2.2.9}
\ee
\quad Furthermore, being $\mathcal G(t_b,t_a)$ trace preserving, we must have
\be
{\rm Tr} \{\mathcal L(t)\hat \rho\}=0 \,.
\lb{2.2.10}
\ee
\\
Under some additional assumption, eq. (\ref {2.2.10}) and the requirement   
$ \mathcal{F}(t_b,t_a;M)$ and so $\mathcal{G}(t_b,t_a)$ be a 
{\it positive mapping} (actually a {\it completely} positive mapping) imply $\mathcal L(t)$ 
to be of the general form
\be
\mathcal L(t) \hat \rho = - \sum_{s=1}^p \alpha_s ( \hat R^{s \dagger} 
\hat R^s \hat \rho + \hat \rho \hat R^{s^\dagger} \hat R^s -2 \hat R^s 
\hat \rho \hat R^{s\dagger})\,.
\lb{2.2.11}
\ee
$\alpha_s$ being appropriate positive constants (cf. \cite{Gorini}).\\

\quad Conversely we can set
\be
\mathcal F(t_b,t_a;\mathcal D_{\rm c} M)= {\bf f} (t_b,t_a;[a(t)])\,\mathcal D_{\rm c} M\,,
 \lb{2.2.12}
\ee
with
\bea
&& {\bf f} (t_b,t_a;[a(t)])= \qquad \qquad \qquad \qquad \nn \\
&& = \int  \mathcal D_{\rm c} \xi \, \exp \left \{i\sum_{s=1}^p \int_{t_a}^{t_b}dt\,
 \xi^s(t)\,a^s(t)\right \} \mathcal G(t_b,t_a;[\xi(t)] \,,
\lb{2.2.13}
\eea
where the measure $\mathcal D_{\rm c} \xi$ is normalized in such a way that
\be
\int  \mathcal D_{\rm c} \xi \,\exp \left\{-i\sum_{s=1}^p \int_{t_a}^{t_b}dt\,
\xi ^s(t)\left(a^s(t)-a^{\prime s}  (t)\right)\right \}= \delta([a(t)]-[a'(t)])\,,
\lb{2.2.14}
\ee
$\delta([a(t)]-[a^\prime (t)])$ being the $\delta$ - functional with respect to the 
measure $\mathcal D_{\rm c} a$.

\quad Formally, this may be achieved assuming the interval $(t_a,t_b)$ divided in $N$ equal 
parts of amplitude $\epsilon = (t_b-t_a)/N$, define
\be
\mathcal D_{\rm c} M \equiv \mathcal D_{\rm c} a = \left ({\epsilon \over 2 \pi}\right)^{Np/2}
 d^p a_1  \dots d^p a_N\,,
\qquad  \mathcal D_{\rm c} \xi = \left({\epsilon \over 2 \pi}\right )^{Np/2} d^p \xi_1 
\dots d^p \xi_N\,,
\lb{2.2.15}
\ee
and
\be
\delta \left ([a(t)]-[a^\prime (t)]\right)=\left ({2\pi \over \epsilon} \right )^{Np/2}\,
\delta^p(a_1-a_2^\prime)\dots \delta(a_N-a_n^\prime),
\lb{2.2.16}
\ee
to be understood in the limit $N\to \infty$.

\quad Eqs. (\ref{2.2.12}) and (\ref{2.2.13}) enable us to reconstruct $\mathcal F (t_b, t_a; M)$  
given $\mathcal G (t_b,t_a;[\xi])$ or $\mathcal K (t, \xi(t))$. Naturally, 
$\mathcal K(t,\xi(t))$ has to be of an appropriate form in order that 
${\bf f}(t_1,t_0;[a(t)])$, as defined by (\ref{2.2.13}) be completely positive.

\quad Two such forms are known: the \textit{Gaussian form}, the
\textit{Poissonian form} and obviously a sum of the two.

\textit{Gaussian form}:
\be
\mathcal K(t,\xi(t))\hat \rho =\mathcal L(t)\hat \rho - i\sum_{s=1}^p\xi ^s(t)
(\hat R^s(t)\hat \rho + \hat \rho \hat R^{s^\dagger} (t)) -\sum_{s=1}^p {1\over
  4\alpha _s}\xi^{s\,2}(t)\,\hat \rho
\lb{2.2.17}
\ee

\textit{Poissonian form}:
\be
\mathcal K(t,\xi(t)) \hat \rho =\mathcal L(t) \hat \rho + 
2 \sum_{s=1}^p \alpha_s (e^{-i\xi^s(t)/2 \alpha_s}-1)
(\hat R^s(t)\hat \rho \hat R^{s\dagger} (t))\,. 
\lb{2.2.18}
\ee

\quad Notice that from (\ref{2.2.13}), (\ref{2.2.14}) we have
\be
\int \mathcal D_{\rm c} a \, {\bf f} (t_b,t_a;[a(t)]) = \int \mathcal D _{\rm c}\xi \,
\,\delta([\xi])\, \mathcal G (t_b, t_a;[\xi]) = \mathcal G (t_b,t_a)
\lb{2.2.19}
\ee
and so 
\be
\int \mathcal D_{\rm c} a\, {\rm Tr}\left \{{\bf f} (t_b,t_a;[a(t)]) \hat\rho (t_a)
\right \}= {\rm Tr} \hat\rho (t_a)=1 \,.
\lb{2.2.20}
\ee
Likewise, for the momenta of the components of $a(t)$ at certain definite times 
$t_1,t_2, \dots t_N$ in the interval $(t_a,t_b)$
\bea
\lb{2.2.21}
&& \langle  a^{s_1} (t_1)\, a^{s_2} (t_2) \dots a^{s_l} (t_l) \rangle =
\int D_{\rm c} a \,\, a^{s_1} (t_1)\, a^{s_2} (t_2) \dots a^{s_l} (t_l) \\
&& {\rm Tr}\left \{{\bf f} (t_b,t_a;[a(t)]) \hat\rho (t_a)\right \} 
 = i^l \, {\rm Tr}\left \{{\delta \over \delta \xi^{s_1} (t_1)} \dots
{\delta \over \delta \xi^{s_l} (t_l)} \mathcal G (t_b, t_a;[\xi]) \hat \rho (t_a)
\right \} \Big \vert_{\xi=0}\,.
\nn  
\eea
In particular we have for the expectation value of a single component
\begin{eqnarray}
&& \langle a^s(t)\rangle = {\rm Tr}\left [i\,{\delta \over \delta \xi^s(t)}
\,\,\mathcal G(t_b,\,t_a;\, [\xi])|_{\xi =0}\,\, \hat \rho (t_a) \,\right ] = \\
&&  = {\rm Tr}\left [\mathcal G(t_b,\, t)\, i\,{\partial \over \partial \xi^s}\, 
\mathcal K(t,\xi)|_{\xi =0}\,\mathcal G(t,t_a)\, \hat \rho(t_a) \,\right ]= 
{\rm Tr}[\hat A^s\, \mathcal G(t,t_a) \,\hat \rho(t_a)\,] \nn
\label{2.2.22}
\end{eqnarray}  
with
\be
\hat A^s(t)=\hat R^s(t)+\hat R^{s\dagger} (t)
\lb{2.2.23}
\ee
in the \textit{Gaussian} case and
\be
\hat A^s(t)=\hat R^{s\dagger}(t)\hat R_s (t)
\lb{2.2.24}
\ee
in the  \textit{Poissonian} case.

\quad Therefore we can talk of $a^1 (t), a^2 (t), \dots ,a^p(t)$ as the value at time 
$t$ of the macroscopic or, in our interpretation, the classical counterpart of the 
quantum observable associated to $\hat A^s$ even if $A^1, A^2, \dots, A^p$ do not each 
others commute.

\quad For the second momenta we have
\bea
&& \langle a^s(t)\, a^{s ^\prime} (t^\prime) \rangle = \delta (t- t^\prime )\, {\rm Tr}
\left \{  {\partial ^2 \mathcal K (t,\xi) \over  \partial \xi ^s 
\partial \xi ^ {s'}} \mathcal G (t, t_a) \hat \rho \right \} \Big \vert_{\xi =0}- \nn \\
&& \qquad - \theta (t-t^\prime) \, {\rm Tr}
\left \{  {\partial K (t,\xi) \over \partial \xi ^s } 
\mathcal G (t, t^\prime) {\partial K (t^\prime,\xi) \over \partial \xi ^{s'} }
\mathcal G (t^\prime, t_a)\hat \rho \right \}\Big \vert_{\xi =0}- \nn \\
&& \qquad - \theta (t^\prime-t) \, {\rm Tr}
\left \{  {\partial K (t^\prime,\xi) \over \partial \xi ^{s^\prime} } 
\mathcal G (t^\prime, t) {\partial K (t,\xi) \over \partial \xi ^s }
\mathcal G (t, t_a)\hat \rho \right \}\Big \vert_{\xi =0}\,. 
\lb{2.2.25}
\eea

\quad The occurrence of the $\delta$ term in eq. (\ref{2.2.25}) shows that only 
{\it time averages} of the type   
\be
a_h = \int dt \, h(t)\cdot a(t) \equiv \int dt \sum _{s=1} ^p h^s (t) a^s (t)\,,
\lb {2.2.26}
\ee
are significant, where the weight functions $h(t) \equiv (h^1 (t), \dots, h^p (t))$are 
elements of the dual space $\mathcal Y ^\prime$. Therefore the class $\Sigma$ of the 
measurable set in $\mathcal Y$ should be defined in terms of such quantities 
(cf. the following subsection). More simply, we may refer to density of 
probability of the form
\bea
\lb {2.2.27}
&& p(a_1,\, h_1;\,a_2,\, h_2;\, \dots a_l,\, h_l)= \\
&& = \int \mathcal {D}_{\rm c} a \,\delta (a_1- \bar {a} _{h_1}) \dots 
\delta (a_l- \bar {a} _{h_l}) \, {\rm Tr} \left \{{\bf f}(t_b, t_a;[a])\, \hat\rho \right \}=
 {1 \over (2 \pi)^{l}}\int d k_1 \dots \nn \\
&& \dots d k_l \,e^{i(k_1 a_1+\dots k_l a_l)}
\int \mathcal{D}_{\rm c} a \, e^{-i(k_1 \bar a _{h_1}+\dots \, k_l \bar a _{h_l} )}\,
{\rm Tr} \left \{{\bf f}(t_b, t_a;[a])\, \hat\rho \right \} = \nn \\
&&= {1 \over (2 \pi)^{l}}\int d k_1 \dots d k_l \,e^{i(k_1 a_1 +n\dots k_l a_l)}
{\rm Tr} \left \{\mathcal{ G}(t_b, t_a;[k_1 h_1 +\dots \,k_l h_l]\, 
\hat\rho \right \}\,, \nn 
\eea
where $h_1 (t), h_2(t), \dots\, h_l(t)$ are independent elements of $\mathcal Y$ with 
support in $(t_a,t_b)$. A sensible choice could be 
\be 
\lb{2.2.28}
h_j (t) = \left (n_j^1\, h(t-t_j),\, n_j^2\, h(t-t_j),\, \dots \, n_j^p\, h(t-t_j)\right )\, ,
\ee 
$h(t)$ being a function different from zero only in a narrow neighbouring of  $t=0$ 
such that $\int dt \, h(t) =1$ ; $\, n_j$ for $j=1,2, \dots l$  unitary vectors in the
euclidean $p$ dimensional space ${\bf R} ^p$; $\, t_1, t_2, \dots t_l $ certain intermediate 
times between $t_a$ and $t_b$.

% 2.3 2.3 2.3 2.3 2.3 2.3 2.3 2.3 2.3 2.3 2.3 2.3 2.3 2.3 2.3 2.3 2.3 2.3 2.3 2.3 

\subsection{Characterization of the history space $\mathcal Y$
and of the measurable set class $\Sigma$}

 \quad Typically one can choose $\mathcal Y = \mathcal E^\prime \times \mathcal E^\prime 
\times \dots \mathcal E^\prime$ and $\mathcal Y ^\prime = \mathcal E \times 
\mathcal E \times \dots \mathcal E$, where $\,\mathcal E\,$ is the class of the function with 
compact support in the $t$-axis and infinitely differentiable everywhere with the 
possible exception of finite discontinuities on the border of the support and 
$\mathcal E ^\prime$ the dual space of $\mathcal E$ (a subset of the Schwartz distribution
space; punctual spectrum is not allowed). The possibility of finite discontinuities on 
the border has to be admitted to give full meaning to eq.(\ref{2.2.2}).

\quad Furthermore we recall that, if  $h_1 (t), h_2(t), \dots\, h_l(t)$ are a set of elements 
of $\mathcal Y  ^\prime$ as above and $B$ a Borel set in ${\bf R}^n$, the subset of 
$\mathcal Y$
\be
\lb{2.3.1}
C(h_1 , h_2, \dots\, h_n ; B)=\left \{a(t) \in \mathcal Y ; (\bar a _{h_1},\bar a _{h_1}
\dots \, \bar a _{h_1}) \in B \right \}
\ee
is called a {\it cylinder set}. Then $\Sigma$ can be identified with the $\sigma$-algebra 
generated by all the cylinder sets for any choice of $n$, $B\subset {\bf R}^n$ and of 
$h_1 , h_2, \dots\, h_n $. The sub-algebra $\Sigma_{t_a}^{t_b}$ is the same, but  with $h_j$ 
with support in the interval $(t_a,t_b)$..

%3333333333333333333333333333333333333333333333333333333333333333333333333

\section{Two specific Gaussian examples}

\quad Let us now consider two specific simple examples, the case of a non relativistic 
harmonic oscillator and the case of a relativistic scalar field in which the quantities 
continuously monitored are a {\it macroscopic position} $q(t)$ and a {\it macroscopic field} 
$\varphi (x)$ respectively in the sense of eq. (\ref{2.2.23}).

%3.1 3.1 3.1 3.1 3.1 3.1 3.1 3.1 3.1 3.1 3.1 3.1 

\subsection{Harmonic oscillator}

\quad Let us write the Lagrangian
\be
L={1 \over 2} (\dot Q^2 -\omega^2 Q^2)\,,
\lb{3.1.1}
\ee
the quantum Hamiltonian
\be
\hat H ={1 \over 2}\, (\hat P^2 + \omega ^2 \hat Q^2)\,,
\lb{3.1.2}
\ee
and commutation rules
\be
[\hat  Q(t),  \hat P(t)]=i \,.
\lb{3.1.3}
\ee

\quad Then we assume in (\ref{2.2.11}, \ref{2.2.17})
\be
\hat R(t)=\hat R^\dagger (t)={1 \over 2}\hat Q(t) \,
\lb{3.1.4}
\ee
and consequently 
\be
\mathcal L (t) \hat \rho = -{\alpha
  \over 4}[\hat Q ,[ \hat Q,\,\hat \rho]] 
\lb{3.1.5}
\ee
and
\bea
\mathcal K(t;\xi(t))\hat \rho = -{\alpha \over 4}[\hat Q ,[ \hat Q,\,\hat
      \rho]]- {i\over 2} \xi(t)\{\hat Q ,\hat \rho \}-{1 \over 4\alpha}\xi^2(t)\,,
\lb{3.1.6}
\eea 
$\alpha$ being a positive constant.

\quad In view of the following developments we want reproduce a proof of the positivity of the
corresponding ${\bf f}(t_b,t_a;[q(t)])$ as defined according to (\ref{2.2.13}) using the 
formalism of the path integral \cite{BLP}. Actually, in the perspective of our interpretation, 
we shall refer to the entire interval $(t_F,\,t_0)$, the restriction to a sub-interval being 
then trivial.

\quad Let us divide the interval $(t_F,t_0)$  in $N$ infinitesimal parts with 
$\epsilon = (t_F - t_0)/N$ and  $t_j=t_0+j\epsilon$. Let us also set to simplify notation
\be
\hat \rho_{[\xi]}(t)=\mathcal G (t,t_0;[\xi]) \rho_0\,.
\lb{3.1.7}
\ee
We have
\bea
&& \hat \rho_{[\xi]}(t_{j+1}) = \hat \rho_{[\xi]}(t_j)+ \epsilon \left \{-{\alpha \over 4}
[\hat Q (t_j),[\hat Q (t_j), \hat \rho_{[\xi]}(t_j)]]\right. \nn \\
&& \qquad \qquad \left.
-{i \over 2} \xi(t_j)\left (\hat Q(t_j)\rho_{[\xi]}(t_j) 
+\rho_{[\xi]}(t_j)\hat Q(t_j)\right )
-{\xi^2_j \over 4\alpha} \hat \rho_{[\xi]} (t_j)\right \}
\lb {3.1.8}
\eea
and, denoting by $|Q,t\rangle$ the eigenstates of $\hat Q (t)$,
\bea
\lb {3.1.9} 
&&\langle Q_{j+1}, t_{j+1}|\hat \rho_{[\xi]}(t_{j+1})|Q^\prime_{j+1}, t_{j+1}\rangle = 
=\int dQ_j \int dQ^\prime_j \nn \\ 
&&\qquad \langle Q_{j+1}, t_j+\epsilon |Q_j, t_j \rangle 
\langle Q_{j}, t_{j}|\hat \rho_{[\xi]}(t_{j})|Q^\prime_{j}, t_{j}\rangle
\langle Q^\prime_{j}, t_j |Q^\prime_{j+1}, t_j +\epsilon \rangle \nn\\
&&\qquad \qquad \qquad \left\{1+\epsilon \left (-{\alpha \over 4}(Q_j-Q^\prime_j)^2 -
{i \over 2} \xi_j(Q_j+Q^\prime_j)-{\xi^2_j \over 4\alpha}\right ) \right \} = \nn\\
&& =\int {dQ_j \, dP_j \over 2\pi}\int {dQ^\prime_j \, dP^\prime_j \over 2\pi}
\exp\left\{\left(iP_j(Q_{j+1}-Q_j) -{\epsilon \over 2}(P^2_j +\omega^2
Q^2_j) \right ) \right \}\cdot \nn \\
&&\langle Q_{j}, t_{j}|\hat \rho_{[\xi]}(t_{j})|Q^\prime_{j}, t_{j}\rangle \,
 \exp\left \{- \epsilon \left ({\alpha \over 4}(Q_j-Q^\prime_j)^2 +
{i \over 2} \xi_j(Q_j+Q^\prime_j)+{\xi^2_j \over 4\alpha}\right ) \right \}\nn \\
&& \qquad \qquad \exp\left\{\left(iP^\prime_j(Q^\prime_{j+1}-Q^\prime_j) 
-{\epsilon \over 2}(P^\prime\,^2_j +\omega^2
Q^{\prime \, 2}_j) \right ) \right \}= \nn \\
&&=\int{dQ_j\,dQ^\prime_j \over 2\pi\epsilon}\exp\left \{
- \epsilon \left ({\alpha \over 4}(Q_j-Q^\prime_j)^2 +{i \over 2} \xi_j(Q_j+Q^\prime_j)+
{\xi^2_j \over 4\alpha}\right )\right .\nn \\
&& \qquad \left . +{i \over 2}\left ({1 \over \epsilon} (Q_{j+1}-Q_j)^2 -
\epsilon \omega^2 Q^2_j\right) - {i \over 2}\left ({1 \over \epsilon} 
(Q^\prime_{j+1}-Q^\prime_j)^2 - \epsilon \omega^2 Q^\prime\,^2_j\right)\right \} \cdot \nn \\
&& \qquad \qquad \qquad \qquad \cdot \langle Q_{j}, t_{j}|\hat \rho_{[\xi]}(t_{j})
|Q^\prime_{j}, t_{j}\rangle \, .
\eea
\quad Iterating eq. (\ref{3.1.9}) we obtain finally for the characteristic operator
\bea
\lb{3.1.10}
&& \langle Q_F, t_F| \mathcal G(t_F,\,t_0;\,[\xi])\hat\rho_0 |Q^\prime_F, t_F \rangle \equiv
 \langle Q_F, t_F|\hat \rho_{[\xi]}(t_F)
|Q^\prime_F, t_F\rangle = \nn \\
&& = \int dQ_0 \int dQ^\prime_0 \,\,\langle Q_0,t_0|\hat \rho_0
|Q^\prime t_0 \rangle 
 \int \mathcal D Q \int \mathcal D Q^\prime \\ 
 &&\qquad \qquad \exp \,\sum_{j=0}^{N-1}\left \{-\epsilon\left({\alpha \over 4}(Q_j-Q^\prime_j)^2 +
{i \over 2} \xi_j(Q_j+Q^\prime_j)+
{\xi^2_j \over 4\alpha}\right )+ \right.\nn \\
&&  \left .  +{i \over 2}\left ({1 \over \epsilon} (Q_{j+1}-Q_j)^2 -
\epsilon \omega^2 Q^2_j\right) - {i \over 2}\left ({1 \over \epsilon} 
(Q^\prime_{j+1}-Q^\prime_j)^2 - \epsilon \omega^2 Q^\prime\,^2_j\right)\right \}\,,\qquad \nn
\eea
\\
where we have set $Q_N=Q_F$, $Q^\prime_N =Q^\prime_F$, 
\be
\lb{3.1.11}
\mathcal DQ=\left({1 \over 2\pi \epsilon}\right) ^{N \over 2}\prod_{j=1}^{N-1}dQ_j
\ee
(the usual Feynman measure) and similarly for $\mathcal DQ^\prime$ . 

\quad Having in mind the ideal limit $N \to \infty$, we can also write in the 
continuous notation
\bea
\lb{3.1.12}
&& \langle Q_F, t_F| \mathcal G(t_F,\,t_0;\,[\xi])\hat\rho_0 |Q^\prime_F, t_F \rangle =\\
&&\qquad \qquad =\int dQ_0 \int dQ^\prime _0 \,\,\langle Q_0,t_0|\hat\rho_0|Q^\prime_0 , t_0
\rangle
 \int_{Q_0}^{Q_F} \mathcal D Q \int _{Q_0^\prime}^{Q_F^\prime}\mathcal D Q^\prime \nn \\
&& \exp \,\int_{t_0}^{t_F}dt \left \{-\left({\alpha \over 4}(Q-Q^\prime)^2 +
{i \over 2} \xi(Q+Q^\prime)+{\xi^2 \over 4\alpha}\right )\right .\nn \\
&&  \left .\qquad\qquad \qquad \qquad +{i \over 2}\left ( \dot Q^2 - \omega^2 Q^2\right) 
- {i \over 2}\left (\dot Q^\prime\,^2 - \omega^2 Q^\prime\,^2\right)\right \}\,,\nn
\eea
where by the extremes in the functional integral we mean that the integration is performed 
keeping the initial and final values of $Q(t)$ and $Q^\prime (t)$ fixed, i. e. under the condition 
$Q(t_0)=Q_0\,,\;Q(t_F)=Q_F$ and $Q^\prime(t_0)=Q_0^\prime\,,\;Q^\prime(t_F)=Q_F^\prime$.  
Note that (\ref{3.1.10}) is the path integral equation corresponding to (\ref{2.2.5}).

\quad Let us then calculate the operation ${\bf f }(t_F,t_0;[q(t)])$ according eq. 
(\ref{2.2.13}), 
$q(t)$ being now the macroscopic position of the oscillator 
\bea
\lb{3.1.13}
&&\langle Q_F, t_F|{ \bf f }(t_F,t_0;[q(t)]) \rho_0|Q^\prime_F, t_F \rangle= \\ 
&&\qquad = \int \mathcal D_c \xi \exp \left \{i\int_{t_0}^{t_F} dt \xi(t)\,q(t)\right \} 
\langle Q_F, t_F| \mathcal G(t_F,\,t_0;\,[\xi])\hat\rho_0 |Q^\prime_F, t_F \rangle =
\nn \\
&& = \int dQ_0 \int dQ^\prime_0 \,\,\langle Q_0,t_0|\rho_0|Q^\prime_0,t_0 \rangle
 \int_{Q_0}^{Q_F} \mathcal D Q \int _{Q_0^\prime}^{Q_F^\prime}\mathcal D Q^\prime 
\left ({\epsilon \over 2\pi}\right )^{N-1 \over 2}\int\prod_{l=1}^{N-1}d\xi_l \nn \\
&&\qquad \qquad \exp \,\sum_{j=0}^{N-1}\left \{i\epsilon \xi_jq_j 
-\epsilon\left({\alpha \over 4}(Q_j-Q^\prime_j)^2 +
{i \over 2} \xi_j(Q_j+Q^\prime_j)+
{\xi^2_j \over 4\alpha}\right )+\right .\nn \\
&&  \left. +\,{i \over 2}\left ({1 \over \epsilon} (Q_{j+1}-Q_j)^2 -
\epsilon \omega^2 Q^2_j\right) - {i \over 2}\left ({1 \over \epsilon} 
(Q^\prime_{j+1}-Q^\prime_j)^2 - \epsilon \omega^2 Q^\prime\,^2_j\right)\right \}= \nn \\
&& =(2\alpha)^{N\over2} \int dQ_0 \int dQ^\prime_0 \,\,
\langle Q_0,t_0|\hat\rho_0|Q^\prime_0 , t_0\rangle
 \int_{Q_0}^{Q_F} \mathcal D Q \int _{Q_0^\prime}^{Q_F^\prime}\mathcal D Q^\prime \nn \\
&&\qquad \qquad \exp\,\sum_{j=0}^{N-1}\left \{-{\alpha \epsilon \over 2}\left ((q_j-Q_j)^2 
+(q_j-Q^\prime _j)^2\right) \right.\nn \\
&&\left . +\,{i \over 2}\left ({1 \over \epsilon} (Q_{j+1}-Q_j)^2 -
\epsilon \omega^2 Q^2_j\right) - {i \over 2}\left ({1 \over \epsilon} 
(Q^\prime_{j+1}-Q^\prime_j)^2 - \epsilon \omega^2 Q^\prime\,^2_j\right)\right \} \nn 
\eea
or, in the continuous limit,
\bea
\lb{3.1.14}
&&\langle Q_F, t_F|{ \bf f }(t_F,t_0;[q(t)]) \rho_0|Q^\prime_F, t_F \rangle= \nn \\
&& \qquad =C_\alpha \int dQ_0 \int dQ^\prime_0 \,\,
\langle Q_0,t_0|\rho_0|Q^\prime_0, t_0 \rangle
 \int_{Q_0}^{Q_F} \mathcal D Q \int _{Q_0^\prime}^{Q_F^\prime}\mathcal D Q^\prime \nn \\
&& \exp\,\int_{t_0} ^{t_F} dt\left \{-{\alpha \over 2}\left ((q-Q)^2 
+(q-Q^\prime)^2\right)\right.\nn \\
&&\qquad \qquad \left.  +{i \over 2}\left ( \dot Q^2 -
\omega^2 Q^2\right) - {i \over 2}\left ( 
\dot Q ^{\prime\,2} -  \omega^2 Q^{\prime\,2}\right) \right\}\,,
\eea
where $C_\alpha$ is a normalization constant formally infinite in the limit $N \to \infty $ , 
which may be possibly incorporated in the classic measure ($\mathcal D_c' q =
(2\alpha)^{N \over 2}\mathcal D_c q=({\alpha \epsilon \over \pi})^{N\over 2} dq_0\, dq_1 \dots dq_{N-1}$).

\quad In operator notation we can also write 
\bea
&& { \bf f }(t_F,t_0;[q(t)]) \hat \rho_0 = C_\alpha \,
\, {\rm T} \exp \left [ {-{\alpha \over 2}\,\int_{t_0}^{t_1}dt\,
  (q(t)- \hat Q (t))^2}\,\right ] \hat \rho  \nn \\
&& \qquad \cdot \,{\rm T^\dagger } \exp \left [{-{\alpha \over
	2}\,\int_{t_0}^{t_1} dt\, (q(t)- \hat Q(t))^2} \right ] \,.
\lb{3.1.15}
\eea
\quad Eq. (\ref{3.1.14}) or (\ref{3.1.15}) shows that ${ \bf f }(t_F,t_0;[q(t)])$ is a 
(completely) positive mapping.

\quad Notice that eqs. (\ref{2.2.22}) and (\ref{2.2.25}) become
\be
\langle q(t) \rangle = {\rm Tr} \left \{ \hat Q (t) \mathcal {G} (t,t_0) \hat \rho _0
\right \}
\lb {3.1.16}
\ee   
and
\bea 
\lb{3.1.17}
&&\langle q(t) q(t^\prime) \rangle = {1 \over 2 \alpha} \delta (t-t^\prime)  \\
&&\qquad \qquad + \,\theta (t-t^\prime) {1\over 2} {\rm Tr}\left ( \hat Q(t)\, 
\mathcal G (t,t^\prime)
\{\hat Q(t^\prime),\, \mathcal G (t^\prime,\, t_0)\hat \rho_0 \} \right ) \nn \\
&& \qquad \qquad + \, \theta (t^\prime-t) {1\over 2} {\rm Tr}\left ( \hat Q(t^\prime)\, 
\mathcal G (t^\prime ,t)
\{\hat Q(t),\, \mathcal G (t,\, t_0)\hat \rho_0 \} \right ) \,.\nn
\eea

\quad Then, let us introduce the quantity (cf. eq. (\ref{2.2.26}))
\be
\lb{3.1.18}
q_h (t) =\int dt^{\prime \prime}\, h(t - t^{\prime \prime})\, q(t^{\prime \prime})
\ee
and set
\be
\lb{3.1.18a}
 \langle \hat A (t) \rangle_{\rm QM} 
= {\rm Tr}\left (\hat A (t)\mathcal G (t,t_0) \hat \rho _0 \right )\,,
\ee
as the most direct analogous quantity to the ordinary Quantum Mechanics expectation value.
Typical assumptions for $h(t)$ may be 
\be
\lb {3.1.19}
 h(t) = {1 \over \tau} \chi _{(-{\tau \over 2},\,{\tau \over 2})} (t)\,,\qquad  
{1 \over \tau \sqrt \pi} e^{-{t^2 \over \tau ^2}} \,,
\ee
$\chi _{(-{\tau \over 2},\,{\tau \over 2})} (t)$ being the characteristic function of the 
interval $(-{\tau \over 2},\,{\tau \over 2})$ . 

\quad We have 
\be
\lb{3.1.20}
\langle q_h (t) \rangle = \langle \hat Q_h (t) \rangle _{\rm QM}
\ee
and, e. g. for the first choice,  
\be
\lb{3.1.21}
\langle\left (q_h(t)-\langle q_h (t) \rangle \right )^2 \rangle ={1 \over 2\alpha \tau} 
+ \left\langle \left (\hat Q _h (t) - \langle \hat Q _h (t) \rangle_{\rm QM} \right )^2 
\right\rangle _{\rm QM}\,,
\ee
in case  $\tau$ is so small that  $\mathcal G (t,t^\prime)$ can be 
replaced by the identity for $t-t^\prime<\tau $. 

\quad The first term in (\ref{3.1.17}) or (\ref{3.1.21}) has no counterpart in ordinary 
quantum theory and is what we have called the {\it intrinsic} part of the variance is the 
introduction.

%  3.2 3.2 3.2 3.2 3.2 3.2 3.2 3.2 3.2 3.2 3.2 3.2 3.2 3.2 3.2 3.2 3.2 3.2 3.2 3.2

\subsection{Real scalar field}
\quad 
To extend the formalism of sect. 2 to fields and make the theory relativistic covariant, 
eq. (\ref{2.2.4}) has to be replaced by 
\be
{\delta \over \delta \sigma (x)}\mathcal G(\sigma , \sigma_0; [j(x)])
= \mathcal K(x,\,j(x)) \mathcal G(\sigma , \sigma_0; [j(x)])\,,
\lb{3.2.1}
\ee
where $x \equiv (t, {\bf x}) $, $\sigma$ and $\sigma_0$ are \textit{spacelike} hypersurfaces and
$\mathcal K(x,\,j(x))$ is expressed only in terms of field operators in the
point $x$.

\quad Note that, in order the above equation to be consistent, the following condition has 
to be satisfied on $\sigma$
\be
\mathcal K(x,\,j(x))\mathcal K(x^\prime,\,j(x^\prime)) = \mathcal
K(x^\prime,\,j(x^\prime))\mathcal K(x,\,j(x))\,.
\lb{3.2.2}
\ee 
Furthermore, identifying $\sigma$ and $\sigma_0$ with time constant hyperplanes and
integrating over the space coordinates, eq. (\ref{3.2.1}) becomes
\be
{\partial \over \partial t}\mathcal G(t, t_0; [j ])
= \int d^3 {\bf x} \, \mathcal K(t,{\bf x};\,j(x)) \mathcal G(t ,
t_0; [j ])\,, 
\lb{3.2.3}
\ee
which is of the general form (\ref{2.2.4}) with ${\bf x}$ playing the role of a 
component index (cf. eq. (\ref{2.2.17}).

\quad Then let us consider the case of a free real scalar field with density of Lagrangian
\be
L(x)={1 \over 2} (\partial _\mu \varphi (x) \, \partial ^\mu \varphi (x) -
m^2\, \varphi^2 (x)) \,. 
\lb{3.2.4}
\ee 
The canonical equal time commutation relation are 
\be
\lb{3.2.5}
[\hat \varphi (x), \hat \varphi (x^\prime)]_{\rm ET} =0, \qquad  
[\hat \varphi (x), \hat \pi (x^\prime)]_{\rm ET} =i \delta (\vec x-\vec x^\prime),
\qquad [\hat \pi (x), \hat \pi (x^\prime)]_{\rm ET} =0 \,,
\ee
with $\hat \pi (x) = \hat \partial _0 \varphi (x)$ and the energy-momentum tensor
\be
\hat T_{\mu\,\nu}(x) = \partial _\mu \hat \varphi (x) \, \partial _\nu \hat \varphi (x) - 
g_{\mu\,\nu}\hat  L(x)\,.
\lb{3.2.6}
\ee

\quad The {\it macroscopic} field $\phi (x)$ is introduced by setting
\be
\mathcal L(x)\, \hat \rho = -{\alpha \over 4 } [\hat \varphi (x) ,
  [\hat \varphi (x), \hat \rho ]]
\lb{3.2.7}
\ee
and
\be
\mathcal K(x,\, j (x))\, \hat \rho =\mathcal L(x)\, \hat
\rho - {i \over 2} j(x)\, \{\hat \varphi (x) , \hat \rho \} -{1 \over 4
    \alpha} j^2(x)\hat \rho \,.
\lb{3.2.8}
\ee

\quad It can be checked that the consistency condition is satisfied if
\be
[\hat \varphi(x),\, \hat \varphi (x^\prime)]=0 \,, \quad  {\rm for \ }  x'  {\ \rm out \  of
  \ the \ light \ cone \ of \ } x \,,
\lb{3.2.9}
\ee
what of course follows from eq. (\ref{3.2.5}) and Lorentz invariance.

\quad Eq.(\ref{3.2.1}) or (\ref{3.2.3}) can be integrated in terms of a path integral as in the 
case of the harmonic oscillator and completely similar developments can be performed. The main 
difference is that now the space-time region between  the initial and 
final hypersurfaces $\sigma_0$ and $\sigma_F$ has to be divided in four-dimensional cells 
of side $\epsilon$ (possibly restricting initially the three-dimensional space to a finite 
volume $V$ and then considering the limit $\epsilon \to 0$ and $V \to \infty$) and 
correspondingly the time integration has to be replaced by a space-time integration.

\quad So instead of (\ref{3.1.12}) we have
\bea
\lb{3.2.10}
&& \langle \varphi_F, \sigma_F| \mathcal G(\sigma_F,\,\sigma_0;\,[j])\hat\rho_0 
|\varphi^\prime_F, \sigma_F \rangle =\\
&&\qquad \qquad \int \mathcal D _{\sigma_0} \varphi_0 \int \mathcal D_{\sigma_0}
\varphi _{\sigma_0}^\prime  \,\,
\langle \varphi_0,\sigma_0|\rho_0|\varphi_0^\prime , \sigma_0
\rangle
 \int_{\varphi_0}^{\varphi_F} \mathcal D \varphi \int _{\varphi_0^\prime}^{\varphi_F^\prime}
\mathcal D \varphi^\prime \nn \\
&& \exp \,\int_{\sigma_0}^{\sigma_F}d^4x \left \{-\left({\alpha \over 4}
(\varphi-\varphi^\prime)^2 +
{i \over 2} j(\varphi+\varphi^\prime)+{j^2 \over 4\alpha}\right )\right .\nn \\
&&  \left .\qquad\qquad \qquad +{i \over 2}\left ( \partial_\mu \varphi\, \partial^\mu
\varphi - m^2 \varphi^2\right) 
- {i \over 2}\left ( \partial_\mu \varphi^\prime\, \partial^\mu
\varphi ^\prime- m^2 \varphi^{\prime\,2}\right) 
\right \}\,,\nn   
\eea
where $|\varphi, \sigma \rangle$ are the simultaneous eigenstates of 
$\hat \varphi (x)$ for $x$ on the spacelike hypersurface $\sigma$ ,
\be
\lb{3.2.11}
\hat \varphi (x) |\varphi, \sigma \rangle = \varphi(x)|\varphi, \sigma \rangle
\qquad  {\rm for \ any }  \qquad x \in \sigma \, , 
\ee
$\mathcal D \varphi$ denotes the functional measure in the space of the $\varphi (x)$s 
regarded as functions in the four dimensional space and $\mathcal D _\sigma \varphi$
the analogous measure for $\varphi (x)$ regarded as functions on the three-dimensional 
surface $\sigma$, i.e. (for simplicity specifically referring to the case in which $\sigma_0$ and 
$\sigma_F$ are equal time hyperplanes, $t=t_0$ and $t=t_F$) 
\be
\lb{3.2.12}
\mathcal D \varphi \simeq \left ({1 \over 2 \pi \epsilon \delta^3}\right )^{N/2}\,
\prod_j\,
d \varphi (x_j) \qquad \mathcal D_\sigma \varphi \simeq  
\left ({1 \over 2 \pi \delta^3}\right)^{N_\sigma/2}\prod_{x_j \in \sigma}
d \varphi (x_j)\,,
\ee
$N$ and $N_\sigma$  being the total number of cells in which the spacetime region delimited
by the volume $V$ and the surfaces  $\sigma_0$ and $\sigma_F$ is divided and the number of cells
intersected  by the surface $sigma$, respectively, $\epsilon$ and $\delta$  the time and 
the space side of each cell.

\quad We have also instead of (\ref{3.1.15})
\bea
&& { \bf f }(\sigma_F,\sigma_0;[\varphi(x)]) \hat \rho_0 = C_\alpha \,
\, {\rm T} \exp \left [ -{\alpha \over 2}\,\int_{\sigma_0}^{\sigma_F}d^4x\,
  (\phi(x)- \hat \varphi(x))^2\,\right ]\, \cdot \nn \\ 
&& \qquad \qquad \cdot \,\hat \rho_0\,{\rm T^\dagger } \exp \left [-{\alpha \over
	2}\,\int_{\sigma_0}^{\sigma_F} d^4x\, (\phi(x)- \hat \varphi(x))^2 \right ] \,.
\lb{3.2.13}
\eea
and 
\bea
\lb{3.2.14}
&&\langle \phi(x)\rangle = -i {\delta \over \delta j(x)}  {\rm Tr}\left [
\mathcal G (\sigma_F ,\sigma_0; [j] ) \hat \rho_0 \right ]|_{j=0}= \nn \\
&&  \qquad \qquad \qquad = {\rm Tr}\left [
\mathcal G (\sigma ,\sigma_0 ) \hat \rho_0 \right ]=
\langle \hat \varphi (x) \rangle _{\rm QM}\, ,
\eea 
$\sigma$ being any spacelike surface trough $x$, and furthermore 
\bea 
\lb{3.2.15}
&&\langle \phi(x) \phi(x^\prime) \rangle = {1 \over 2 \alpha} \delta^4 (x-x^\prime)  \\
&&\qquad \qquad + \theta (t-t^\prime)\, {1\over 2}\, {\rm Tr}\left [ \hat \varphi(x)\, 
\mathcal G (\sigma,\sigma^\prime)
\{\varphi(x^\prime),\, \mathcal G (\sigma^\prime,\, \sigma_0)\hat \rho_0 \} \right ] \nn \\
&& \qquad \qquad + \theta (t^\prime-t)\, {1\over 2}\, 
{\rm Tr}\left [ \hat \varphi(x^\prime)\, 
\mathcal G (\sigma^\prime ,\sigma)
\{\varphi (x),\, \mathcal G (\sigma,\, \sigma_0)\hat \rho_0 \} \right ] \,.\nn
\eea

\quad In place of eq. (\ref{3.1.18}) we have to consider spacetime averages of the type
\be
\lb{3.2.16}
\phi_h (x) = \int d^4 x^\prime h(x^\prime - x) \phi (x^\prime) \,,
\ee
where for instance
\be
\lb{3.2.17}
h(x)={1 \over a^4} \chi _\omega (x) \,,
\ee
$\omega$ being a four-dimensional cube of side $a$ centered on the origin, 
and for $a$ sufficiently small 
\be
\lb{3.2.18}
\left \langle \left ( \phi_h (x) -\langle \phi_h (x) \rangle \right )^2 \right \rangle =
{1 \over 2 \alpha a^4}+\left \langle \left (\hat \varphi_h (x) 
- \langle \hat \varphi _h (x)\rangle
 _ {\rm QM}\right )^2 \right \rangle _{\rm QM}\,.
\ee

%44444444444444444444444444444444444444444444444444444444444444444

\section{Problems with the conservation laws}

\quad At this point we may note that equations of the type (\ref{3.1.5}) and (\ref{3.2.7}) are 
in conflict with the law of energy conservation, what does not seems acceptable in the 
prospective we are pursuing of a modification of ordinary Quantum Theory.

\quad In the present framework, for energy conservation we intend the conservation of 
$\langle \hat H \rangle_{\rm QM}$ 

\qquad In fact for any given observable one has
\bea
&& {d \over dt}\langle \hat A(t) \rangle _{\rm QM}={\rm Tr}\left [{d
  \hat A \over dt}\, \mathcal G(t,t_0)\, \hat \rho \right] 
+ {\rm Tr}\left [ \hat A(t) {\partial \over \partial t}\, 
\mathcal G(t,t_0) \hat \rho \right] = \nn \\
&&\qquad  =\langle {d \hat A(t) \over dt} \rangle _{\rm QM}
+ {\rm Tr}\left [\hat A(t)\, \mathcal L(t) \,  \mathcal G(t,t_0) \hat \rho
\right] = \nn \\
&& \qquad = \langle {d \hat A(t) \over dt} \rangle _{\rm QM}
+ {\rm Tr}\left [(\mathcal L^\prime (t)\hat A(t))\, \mathcal  G(t,t_0) \hat
\rho \right]\,, 
\lb{4.1}
\eea
i. e.
\be
{d \over dt}\langle \hat A(t) \rangle _{\rm QM}=
\langle {d \hat A(t) \over dt} + \mathcal L^\prime (t)\hat A(t)
\rangle _{\rm QM} \,,
\lb{4.2}
\ee
where by $\mathcal L ^\prime $ we denote the dual mapping of $\mathcal L$ (see (\ref{2.4a})).

\quad For the harmonic oscillator we have 
\be
\mathcal L^\prime (t)\hat A(t) = - {\alpha \over 4} [[\hat A(t),\hat Q(t)],
  \hat Q(t)]=- {\alpha \over 4} [\hat Q(t)[ \hat Q(t),\hat A(t)]]\,.\qquad
\lb{4.4}
\ee
Consequently, since obviously ${d  \over dt}\hat H =0 $,
\be
  {d \over dt}\langle \hat H  \rangle _{\rm QM}=\langle \mathcal
L^\prime (t) \hat H \rangle _{\rm QM}={\alpha \over 4}\not= 0
\lb{4.5}
\ee
and the energy is no longer conserved. 

\quad In a similar way for the real scalar field 
\bea
&& \partial _\mu \langle \hat T^{\mu \,\nu}(t, {\bf x})\rangle _{\rm
  QM}= \nn \\
&& = \langle \partial _\mu \hat T^{\mu \, \nu}(t, {\bf x}) + \int d^3
\vec y \, \mathcal  L^\prime (t, {\bf  y})\,\hat T^{0 \, \nu}(t, {\bf x})
\rangle _{\rm QM}={1 \over 4}g^{0\,\nu}\, \delta ({\bf 0}) \,,
\lb{4.6}
\eea
which is not only different from 0 but even infinite and non-covariant, showing that 
$\mathcal L(x)$ as given by (\ref{3.2.7}) is hill defined.

\quad \textit {At a formal level} the above difficulty can be overcome
modifying the definition of $\mathcal L (t)$ and $\mathcal L (x)$.

\qquad For the harmonic oscillator we can replace (\ref{3.1.5}) with 
\bea
\lb{4.7} 
&& \mathcal L(t) = - {\gamma \over 4} \left ([\hat P(t),[
      \hat P(t), \hat \rho]] - \omega ^2 [\hat Q(t), [\hat Q(t),
      \hat \rho]]\right )  \\
&&\qquad \qquad = - {\gamma \over 4} \left ([\dot {\hat Q}(t),[\dot
      {\hat Q}(t), \hat \rho]] - \omega ^2 [\hat Q(t), [\hat Q(t),
      \hat \rho]]\right ) \,, \nn 
\eea
$\gamma$ being now an adimensional constant. Then we have immediately
\be
\mathcal L^\prime (t) \hat H =0 \qquad \Rightarrow \qquad {d \over dt}\,
\langle \hat H \rangle _ {\rm QM}=0 \,.
\lb{4.8}
\ee 

\quad Similarly for the scalar field we can take
\be 
\mathcal L(x)\,\hat \rho =- {\gamma \over 4} ([\partial _\mu \hat \varphi (x),
[\partial ^\mu \hat \varphi (x), \hat \rho]] - m ^2 [\hat \varphi (x),
[\hat \varphi (x), \hat \rho]])\,;
\lb{4.9}
\ee
and then again
\be
\mathcal L^\prime (t,{\bf y})\, \hat T^{0\, \nu}(t, {\bf x}) =0 \qquad \Rightarrow \qquad
\partial _\mu \langle \hat T^{\mu \, \nu}(x) \rangle _ {\rm QM}=0 \,.
\lb{4.10}
\ee

\quad Note the similarity between the factors multiplying $-{\gamma \over
  4}$ in (\ref{4.7}) and (\ref{4.9}) and the corresponding Lagrangian and density of 
Lagrangian (\ref{3.1.1}) and (\ref{3.2.6}) respectively.

\quad Note also  that with the definitions  (\ref{4.7}) and (\ref{4.9}) 
the mappings  $\mathcal G(t,t_0)$ and $\mathcal G(\sigma, \sigma _0)$ are no longer 
positive.
E.  g. we can write
\bea
\lb{4.11}
&&\mathcal G(t +\epsilon, t)\hat \rho = \hat \rho - {\epsilon \gamma \over 4} 
([\hat P(t),[ \hat P(t), \hat \rho]] - \omega ^2 [\hat Q(t), [\hat Q(t),\hat \rho]])= \\
&& = \{ 1 - {\epsilon \gamma \over 4} (\hat P^2 - \omega^2 \hat Q^2 \} \{\hat \rho + 
{\epsilon \gamma \over 2} (\hat P \hat \rho \hat P -\omega ^2 \hat Q \hat \rho \hat Q)\}
 \{ 1 - {\epsilon \gamma \over 4} (\hat P^2 - \omega^2 \hat Q^2) \}\nn
\eea
and in the last line the middle factor is not positive in general.

\quad However we shall show that the definition of $\mathcal K(t, [\xi] )$ and 
$\mathcal K (x,[j])$ can be modified in a corresponding way that the densities
${\bf f}(t_F,t_0; [q])$ and ${\bf f}(\sigma_F,\sigma_0; [\phi])$ remain positive.

%555555555555555555555555555555555555555555555555555555555555555555555555555555

\section{Modified formalism}

%515151515151515151515151515151515151515151515151515151515

\subsection{Harmonic oscillator}

\quad According to (\ref{4.7}) we assume
\be
\mathcal L(t) \hat \rho =- {\gamma \over 4} ([\hat P(t),
[ \hat P(t),\hat \rho ]] - \omega ^2 [\hat Q(t), [\hat Q(t), \rho]])
               \lb{5.1.1}
\ee  
and correspondingly take
 \bea
&& \mathcal K(t,\xi)\hat \rho =  \mathcal L(t) \hat  - {i \over 2} \xi(t) \{ \hat Q(t), \hat \rho \, \}  \nn \\
&& \qquad \qquad - {1  \over 2 \gamma} \int _{t_0} ^t dt^\prime \, G (t-t^\prime)
\,\xi(t)\,\xi(t^\prime)\,\hat \rho \,, 
\lb{5.1.2}
\eea
where
\be
 G (t- t^\prime)={1 \over T }\sum _{n=-\infty} ^\infty{1 \over k_n^2-\omega^2}\, e^{i k_n(t- t^\prime)}
\simeq {1 \over 2 \pi}\int_{-\infty}^\infty dk \,{\rm P}{1 \over k^2-\omega^2}\, 
 e^{i k(t- t^\prime)}
\lb{5.1.3}
\ee
($T=t_F-t_0,\,\,\,\,k_n={2 \pi n\over T}, \,\,\,\, n=0,\,\pm 1,\,\pm 2\,,\dots$)
is the solution of the equation
\be
{\rm K}\, G (t-t^\prime)\equiv
-({d^2 \over dt^2}+ \omega^2)\, G (t-t^\prime)=
\delta (t-t^\prime)
\lb{5.1.4} 
\ee
i.e. it is the inverse of the differential operator ${\rm K}$ .
\footnote {In the expression as an integral in eq. (\ref{5.1.3}) some kind of 
regularization has to be assumed. At the discrete level this amounts to make a choice 
for the ratio between $T$ and the classic period ${2\pi \over \omega}$, in order to avoid 
that some $k_n$ coincides with $\pm \omega$. The principal value 
prescription obviously corresponds to assume ${\omega T \over 2\pi}$ to be an half odd 
integer. However, since actually,as we shall see, we can restrict ourselves to values of 
$|k_n|>\omega$ the specific choice turns out to be irrelevant}  
Obviously we can also write
\bea
&& K (t-t^\prime)\equiv {\rm K} \delta(t-t^\prime)= \nn \\
&& \quad {1 \over T }\sum _{n=-\infty} ^\infty (k_n^2-\omega^2)\, e^{i k_n(t- t^\prime)} 
\simeq {1 \over 2 \pi }\int_{-\infty}^\infty dk\, (k^2-\omega^2)\, e^{ik(t-t^\prime)}\,.
\lb{5.1.5}
\eea

\quad Note that, with our modified definition, $\mathcal K(t,\xi)$ depends not only on 
the value of $\xi$ at the time $t$, as in (\ref{3.1.8}), but on its the entire history 
before $t$. Even with $\mathcal G (t_b,\,t_a;\,[\xi]))$  defined as in (\ref{2.2.5}), 
therefore, it has to be equally understood that the integral in $t'$ is extended from $t_0$ to 
$t$. Then eq. (\ref{2.2.2}) and consequently 
\begin{equation}
\mathcal G(t_c,t_b)\,\mathcal G(t_b,t_a)=
\mathcal G(t_c,t_a) 
\label{5.1.6}
\end{equation}
remain formally valid. On the contrary, strictly speaking Eq. (\ref{2.1.3})
is no longer an exact equation; however, as we shall see, it remains valid as a good approximation 
for all practical purposes.

\quad According (\ref{5.1.1}, \ref{5.1.2}) now instead of (\ref{3.1.8}) we have
\bea
\lb{5.1.7}
&&\hat \rho_{[\xi]}(t_{i+1}) = \hat \rho_{[\xi]}(t_i) -  \\
&& - \epsilon \left \{{ \gamma \over 4} 
\left ([\hat P(t_i),[ \hat P(t_i), \hat \rho_{[\xi]}(t_i)]] - 
\omega ^2 [\hat Q(t_i), [\hat Q(t_i),\hat \rho_{[\xi]}(t_i)]] \right )+ \right. \nn \\
&&\qquad \left .  + {i \over 2} \xi (t)\{ \hat Q (t_i),\,\hat \rho_{[\xi]}(t_i)\}
+{1 \over 2 \gamma}
\int _{t_0}^{t_i} dt^\prime G(t_i- t^\prime)\,\xi(t_i) \xi (t^\prime)\,
\hat \rho_{[\xi]}(t_i) \right \} \nn
\eea
and, by appropriately inserting completenesses in the momentum eigenvectors,
\bea
\lb{5.1.8}
&& \langle Q_{i+1},\,t_{i+1}|\hat\rho_{[\xi]}(t_{i+1}) |Q_{i+1}^\prime,\,t_{i+1}\rangle = \nn \\
&&\qquad = \int {dQ_i \, d P_i \over 2 \pi}\int {dQ_i^\prime \, d P_i^\prime \over 2 \pi}
\exp \left \{i \left [P_i (Q_{i+1}-Q_i) -{\epsilon \over 2}(P_i^2 -\omega^2 Q_i^2)\right ]
\right \}\nn \\
&& \qquad \qquad \qquad \exp \left \{-{\epsilon \gamma \over 4} \left [ (P_i-P_i^\prime)^2-\omega^2 
(Q_i-Q_i^\prime)^2 \right ] - \right. \nn \\
&&\left. \qquad \qquad -{i \epsilon \over 2}\xi_i (Q_i+Q_i^\prime )-
{\epsilon^2 \over 2 \gamma}\sum _{j=0}^i{}'\, G_{i\,j}\, \xi_i \xi_j \right \} \nn \\
&& \exp \left \{i\left [P_i^\prime (Q_{i+1}^\prime-Q_i^\prime)  
-{\epsilon \over 2}(P_i^{\prime \, 2} -\omega^2 Q_i^{\prime\,2})\right ]\right \} \langle
Q_i , \, t_i | \rho_{[\xi]} |Q_i^\prime,\,t_{j}\rangle =  \nn \\
&&  = {1 \over 2 \pi \epsilon} \int dQ_i \int dQ_i^\prime \exp \left \{ -{\gamma \over 4}
\left [{1 \over \epsilon }\left ((Q_{i+1}-Q_i )- \right. \right. \right. \nn\\
&& \qquad \qquad \left. \left. -(Q_{i+1}^\prime-Q_i^\prime ) \right )^2 
 - \epsilon \omega^2 (Q_i-Q_i^\prime)^2 \right ] + \\
&& + {i \over 2} \left ({1 \over \epsilon} (Q_{i+1}-Q_i)^2 - \epsilon \,\omega ^2
Q_i^2 \right )-{i \over 2} \left ({1 \over \epsilon} (Q_{i+1}^\prime -Q_i^\prime)^2
 - \epsilon \omega ^2 Q_i^{\prime\,2} \right ) - \nn \\
&& \left. \qquad \qquad  - {i \epsilon \over 2}\xi_i (Q_i+Q_i^\prime )-
{\epsilon^2 \over 2 \gamma}\sum _{j=0}^i {}' \, G_{ij}\,\xi_i \xi_j \right \}
\langle Q_i , \, t_i | \hat\rho_{G[\xi]}(t_i) |Q_i^\prime,\,t_i\rangle \,, \nn
\eea
where $G_{ij}$ stays for an appropriate discretization of $G(t-t^\prime)$, to be specified 
later, and the prime in the sum indicates that the diagonal elements $G_{ii}$ must to be 
multiplied for ${1 \over 2}$ so that by exploiting the symmetry of such quantity we can write 
$\sum_{i=0}^{N-1}\, \sum_{j=0}^{i \,\prime} \,G_{i\,j} \, \xi_i\, \xi_j= 
{1 \over 2} \sum_{i=0}^{N-1}\,\sum_{j=0}^{N-1} \,G_{i\,j}\xi_i \, \xi_j $.

\quad Iterating eq. (\ref{5.1.8}) over the entire interval $(t_0,\,t_F)$ with the same 
definitions as in (\ref{3.1.11}), we obtain 
\bea
\lb{5.1.9}
&& \langle Q_F, t_F| \mathcal G(t_F,\,t_0;\,[\xi])\hat\rho_0 |Q^\prime_F, t_F \rangle = \nn \\
&& = \int dQ_0 \int dQ^\prime_0 \,\,\langle Q_0,t_0|\hat\rho_0|Q^\prime t_0 \rangle
 \int_{Q_0}^{Q_F} \mathcal D Q \int _{Q_0^\prime} ^{Q_F^\prime} \mathcal D Q^\prime \nn \\
&&\exp \sum_{i=0}^{N-1}\left \{ -{\gamma \over 4}
\left [{1 \over \epsilon }\left ((Q_{i+1}-Q_i )-(Q_{i+1}^\prime-Q_i^\prime ) \right )^2 -
\epsilon \omega^2 (Q_i-Q_i^\prime)^2 \right ]\right . \nn \\
&& \qquad \qquad + {i \over 2} \left ({1 \over \epsilon} (Q_{i+1}-Q_i)^2 - \epsilon \omega ^2
Q_i^2 \right )-{i \over 2} \left ({1 \over \epsilon} (Q_{i+1}^\prime -Q_i^\prime)^2
 - \epsilon \omega ^2 Q_i^{\prime\,2} \right ) \nn \\
&& \left . \qquad \qquad \qquad - {i \epsilon \over 2}\xi_i (Q_i+Q_i^\prime )-
{\epsilon^2 \over 2 \gamma}\sum _{j=0}^i {}' \, G_{i\,j}\xi_i \xi_j \right \}= \nn \\
&& = \int dQ_0 \int dQ^\prime_0 \,\,\langle Q_0,t_0|\hat\rho_0|Q^\prime t_0 \rangle
 \int_{Q_0}^{Q_F} \mathcal D Q \int _{Q_0^\prime} ^{Q_F^\prime} \mathcal D Q^\prime \nn \\
&&\exp \left \{\sum _{i,j=0}^{N} {1 \over \epsilon} K_{i\,j}\left [-{\gamma \over 4}
(Q_i-Q_i^\prime) (Q_j-Q_j^\prime )+ {i \over 2}Q_i Q_j - 
{i \over 2}Q_i^\prime Q_j^\prime \right ] \right .\nn \\
&& \left . \qquad \qquad  - {i \epsilon \over 2} \sum _{j=0} ^ {N-1} 
\xi_j (Q_j+Q_j^\prime )-{\epsilon^2 \over 4 \gamma}
\sum _{i,j=0}^{N-1}\xi_i G_{i\,j} \xi_j \right \} \,,  
\eea
where we have set again $Q_N=Q_F$ and $Q_N'=Q_F'$ and denoted by $K_{i\,j}$ the 
$(N+1)\times (N+1)$ matrix
\be
\lb{5.1.10} 
K_{i\,j}
=\left(\matrix {  1-\epsilon^2 \omega ^2   & -1 & 0 &\dots & 0 & 0 \cr
        -1 & 2-\epsilon^2 \omega ^2 & -1 & \dots & 0 & 0  \cr
0 & -1 &  2-\epsilon^2 \omega ^2 & \dots & 0 & 0  \cr
\dots & \dots & \dots & \dots & \dots & \dots  \cr
0 & 0 & 0 & \dots & 2-\epsilon^2 \omega ^2 & -1 \cr
0 & 0 & 0  &  \dots &  -1  &  1 \cr} 
 \right )\,,
\ee
which obviously provides a discretization of (\ref{5.1.5}).

\quad Now, let us denote by $\bar K_{ij}$ the $N\times N $ matrix 
obtained by suppressing the last row and column in $K_{ij}$. For $N$ sufficiently 
large ($\epsilon$ small)  $\bar K_{ij}$ turns out to be positive (App. B) Furthermore, while 
$\det K_{ij}=0|_{\epsilon =0}$, it is $\det \bar K_{ij}|_{\epsilon =0} =1$. Then we can  chose 
$G_{ij}=\epsilon \bar K_{ij}^{-1}$, which is also a positive matrix, and we have 
\bea
\lb{5.1.11}
&&\left ({\epsilon \over 2 \pi} \right )^{N \over 2} \int d\xi_0 \dots d\xi_{N-1} \,
\exp \left \{i \epsilon \sum_{j=0}^{N-1} \xi_j\,\left (q_j - 
  {Q_j+Q_j^\prime \over 2 }\right )-{\epsilon \over 4\gamma}\sum_{i,j=0}^{N-1}\xi_i  
G_{ij}\xi_j \right\}
\nn \\
&& = \left ({2 \gamma \over \epsilon ^2}\right)^{N\over 2} 
\exp \left \{-{\gamma \over \epsilon } \sum_{i,j=0}^{N-2} 
\left(q_i -{Q_i+Q_i^\prime \over 2}\right) 
\bar K_{ij}\left(q_j -{Q_j+Q_j^\prime \over 2}\right)\right \}\,.
\eea 
 Finally, by similar manipulations to those performed in (\ref{3.1.13}), we obtain 
\bea
\lb{5.1.13.0}
&& \langle Q_F, t_F|{\bf f}(t_F,\,t_0;\,[q(t)])\, 
\hat \rho_0 |Q_F^\prime, t_F \rangle  = \left ({2\gamma \over \epsilon^2} \right )^{N \over 2}
 \int dQ_0\,dQ_0^\prime \nn \\
&& \int_{Q_0}^{Q_F} \mathcal{D}Q \,\exp \sum _{j=0}^{N-2} 
\left \{-\,{\gamma \over 2 }\left [
{1 \over \epsilon}\left ((q_{i+1}-Q_{i+1})-(q_i -Q_i)\right )^2 \right. \right. \nn \\
&&\qquad \left. \left. -\,\omega^2 \epsilon (q_i -Q_i)^2 \right ]+ 
{i \over 2} \left[{1 \over \epsilon}(Q_{i+1}- Q_i)^2 
- \omega^2 Q_i^2\right ]
\right \} \langle Q_0|\hat \rho_0 |Q_0^\prime \rangle  \nn \\
&& \int_{Q'_0}^{Q'_F} \mathcal{D}Q' \,\exp \sum _{i=0}^{N-2} 
\left \{-{\gamma \over 2 } \left [
{1 \over \epsilon}\left ((q_{i+1}-Q_{i+1}^\prime)-(q_i -Q_i^\prime)\right )^2 - \right. \right. \nn \\
&& \qquad \left.\left. -\,\omega^2 \epsilon 
(q_i -Q_i^\prime)^2 \right ]+ {i \over 2} \left[{1 \over \epsilon}
(Q_{i+1}^\prime- Q_j^\prime)^2 -\omega^2\epsilon Q_j^{\prime \,2} 
\right] \right \} \\
&& \qquad\exp \left \{-{\gamma \over 4 \epsilon}\,
\left [(Q_F-Q_F')^2-2(Q_F-Q_F')(Q_{N-1}-Q_{N-1}') \right ] \right \}\,.\nn
 \eea

\quad Eq. (\ref{5.1.13.0}) is analogous to eq. (\ref{3.1.13}) for the original formalism apart 
from the occurrence under the integral of the last exponential factor
\be
\label{5.1.13.1}
\exp \left \{-{\gamma \over 4 \epsilon}\,
\left [(Q_F-Q_F')^2-2(Q_F-Q_F')(Q_{N-1}-Q_{N-1}') \right ] \right \}\,.
\ee
Equivalently, by setting
\be
\label{5.1.13.2}
q_N \equiv q_F={Q_F+ Q_F ' \over 2}\,,
\ee
we can omit the factor (\ref{5.1.13.1}) and extend the sum in (\ref{5.1.13.0})  to $N-1$.

\quad The fact that (\ref{5.1.13.1}) or $q_F$ in (\ref{5.1.13.2}) depend on both the  primed and 
unprimed variables prevents us from concluding that ${\bf f}(t_F,\,t_0;\,[q(t)])$ is 
positive. On the other side, when we define the mapping for restrict intervals  
${\bf f}(t_b,\,t_a;\,[q(t)])$, the factor (\ref{5.1.13.1}) is essential for the validity of 
the equation
\be
\label{5.1.13.3}
 {\bf f}(t_c,\,t_a;\,[q(t)])\,=\,{\bf f}(t_c,\,t_b;\,[q(t)])\,{\bf f}(t_b,\,t_a;\,[q(t)]) \,,
 \ee 
corresponding to eq (\ref{2.1.3}). However, for $Q_F = Q_F'$ the factor (\ref{5.1.13.1}) reduces to 1 and 
${\bf f}(t_F,\,t_0;\,[q(t)])$ coincides with $\bar {\bf f}(t_F,\,t_0;\,[q(t)])$, defined by the 
same eq. (\ref{5.1.13.0}) as it stays but omitting the factor (\ref{5.1.13.1}) and, obviously,  
$\bar {\bf f}(t_F,\,t_0;\,[q(t)])$ is positive. Then even the functional probability density
\be
\label{5.1.13.4} 
p\,(t_F,\,t_0;\,[q(t)]) = {\rm Tr} \left \{{\bf f}(t_F,\,t_0;\,[q(t)])\, \hat \rho_0 \right \}=  
{\rm Tr} \left \{\bar {\bf f}(t_F,\,t_0;\,[q(t)]) \hat \rho_0 \right \}
\ee
is positive and this is what matters. 

 \quad As in sec. 3, in the limit $N \to \infty$ eqs. (\ref{5.1.8}) and 
(\ref{5.1.13.0}) can be formally written
\bea
\lb{5.1.14}
&& \langle Q_F, t_F| \mathcal G(t_F,\,t_0;\,[\xi])\hat\rho_0 |Q^\prime_F, t_F \rangle = \nn \\
&& = \int dQ_0 \int dQ^\prime_0 \,\,\langle Q_0,t_0|\hat\rho_0|Q^\prime, t_0 \rangle
 \int_{Q_0}^{Q_F} \mathcal D Q \int _{Q_0^\prime} ^{Q_F^\prime} \mathcal D Q^\prime \nn \\
&&\exp \int_{t_0}^{t_F} dt \left \{ -{\gamma \over 4}
\left [ (\dot Q - \dot Q^\prime )^2 -
 \omega^2 (Q-Q^\prime)^2 \right ]+ \right . \nn \\
&& \qquad \qquad + {i \over 2}  (\dot Q^2 - \omega ^2
Q^2  )-{i \over 2}  (\dot Q^{\prime\,2}
 - \omega ^2 Q^{\prime\,2} )- \nn \\
&& \left. \qquad \qquad - {i \over 2}\xi (Q+Q^\prime )-
{1 \over 4 \gamma}\int_{t_0}^{t_F}dt^\prime \xi(t)\,G(t-t^\prime)\,\xi(t^\prime) \right \}
\eea
and
\bea
\lb{5.1.15}
&& \langle Q_F, t_F|\bar{ \bf f}(t_F,\,t_0;\,[q(t)])\, \hat \rho_0 |Q_F^\prime, t_F \rangle = 
\left ({2\gamma \over \epsilon^2} \right )^{N \over 2}
\int dQ_0dQ_0^\prime \nn \\
&& \int_{Q_0}^{Q_F} \mathcal{D}Q \,\exp  \int_{t_0}^{t_F}dt^\prime
\left \{-{\gamma \over 2 }\left [
(\dot q-\dot Q )^2-\omega^2 (q -Q)^2 \right ]+ \right .\nn \\
&& \left . \qquad \qquad \qquad\qquad + {i \over 2} (\dot Q^2 -\omega^2 Q^2)
\right \} \langle Q_0|\hat \rho_0 |Q_0^\prime \rangle  \nn \\
&& \int_{Q'_0}^{Q'_F} \mathcal{D}Q' \exp \int_{t_0}^{t_F}dt^\prime 
\left \{-{\gamma \over 2 }\left [ (\dot q-\dot Q^\prime)^2-\omega^2  
(q_j -Q_j^\prime)^2 \right ]+ \right .\nn \\
&& \qquad \qquad \qquad \qquad +\left .{i \over 2} (\dot Q^{\prime\, 2} -\omega^2 Q^2) 
\right \}\,,
\eea
or in operator form 
\bea
 && \bar {\bf f}(t_F,\,t_0;\,[q(t)])\,\hat \rho_0 = \\
&&  \qquad  = C_\gamma\,\, {\rm T}\exp \left \{ -{ \gamma \over
  2}\int _{t_0}^{t_F} dt \,[(\dot q(t)- \dot {\hat Q}(t))^2 
-\, \omega^2 (q(t)-\hat Q(t))^2]\right \} \nn \\
&& \qquad \qquad \hat \rho_0 \,\, {\rm T}^\dagger \exp \left \{
    -{\gamma \over 2}\int _{t_0}^{t_F} dt \,[(\dot q(t)- \dot {\hat Q}(t))^2 
-\omega^2 (q(t)-\hat Q(t))^2]\right \}\,. \nn
\lb{5.1.16}
\eea
\quad  Actually eqs. (\ref{5.1.14}-\ref{5.1.16}) requires some comments. 

\quad What grants convergence of the integral in eq. (\ref{5.1.9}), when performed step 
by step, it is the prevalence of the terms in $1 / \epsilon$ in the exponent 
over the terms in $\epsilon$. This circumstance may be clearly 
illustrated on the similar case of the usual Feynman expression of the ordinary amplitude 
$\langle Q_F,\, t_F|Q_0,\,t_0\rangle$ for the harmonic oscillator, which can be calculated 
exactly and remains valid even for an imaginary mass $m=i\mu$ with positive $\mu$ 
(App. A). The positivity of the matrices $\bar K_{ij}$ and 
$G_{ij}=\epsilon \bar K_{ij}^{-1}$ grants that the convergence of the integrals in 
(\ref{5.1.11}) and (\ref{5.1.13.0}) is even absolute. We have
\be
\lb{5.1.19}
\sum_{i,j=1}^{N-1}\xi_i\,G_{ij}\,\xi_j > 0 \,,
\ee   
for any $(\xi_1, \, \xi_2, \, \dots \,\xi_{N-1})$, that in the limit would correspond to
\bea
&&\int_{t_0}^{t_F}dt \int _{t_0}^{t_F} dt' \xi(t)  G(t-t')  \xi(t')= \qquad \nn \\
&&  \qquad = \sum _{k=-\infty}^\infty \tilde \xi _k^*\, { 1\over k^2-\omega^2}\,\tilde \xi _k 
\sim \int _{\-\infty}^\infty dk \, \tilde \xi^*(k)\,{\rm P} {1\over k^2 -\omega^2 }\,
\tilde \xi(k)>0 \,,\quad  
\lb {5.1.20}
\eea 
where  $\tilde \xi_k$ 's  are the Fourier coefficients and $\tilde \xi (k) \sim 
\sqrt {2\pi \over T}\,\tilde \xi_k$\,. the Fourier transform.

\quad  Such circumstances are somewhat surprising.  The kernel 
$ K(t-t')= -({d^2 \over dt^2 }-\omega^2) \delta(t-t')$   
and its inverse $G(t-t')$ are not positive in $L^2(t_0,\, t_F)$ and the meaning of eq, 
(\ref{5.1.14}-\ref{5.1.16}) seems to be questionable. Actually this shows that the specific 
limit procedure by which the functional integrals have been defined and the continuous 
kernel $G(t-t')$ approached is crucial. Specifically it implies that only the Fourier 
components of $Q(t)-Q_c (t)$, $Q' (t)-Q_c' (t)$ and $\xi(t)$ with $|k|>\omega$ give non 
vanishing contribution in the integrals, $Q_c (t)$ and $Q_c' (t)$ denoting the 
solutions of the classical equation of motion satisfying the conditions
\be
                Q_c (t_0 )=Q_0 \, ,  \qquad \qquad \qquad       Q_c (t_F )=Q_F  
\lb {5.1.21}
\ee                          
and
\be
               Q_c'(t_0 )=Q_0' \,, \qquad \qquad \qquad        Q_c'(t_F )=Q_F' \,.
\lb {5.1.22}
\ee
This fact can be also explicit checked by considering the continuous function $Q(t)$ and $Q(t')$ 
obtained by interpolating linearly the discrete values $Q_0,\,Q_1,\,\dots\, Q_F$ and 
$Q_0',\,Q_1',\,\dots\, Q_F'$ respectively (App. C); their components with $|k|<\omega$ vanish
for $\epsilon \to 0$. 

\quad Alternatively, to be able to proceed directly in the more appealing continuous formalism, by 
using only general properties of the functional integral, it is convenient to assume  
explicitly that all the functions of interest, $Q(t)$, $Q'(t)$, $\xi(t)$ (and consequently $q(t)$) 
are restricted to the subspace with Fourier components with $|k|\geq \omega$ and that the boundary 
conditions.
\be
\lb{5.1.22a}
\xi(t_0)=\xi(t_F)=0
\ee
and
\be
\lb{5.1.22b}
q(t_0)= {1 \over 2}(Q_0 + Q_0^\prime ), \qquad \qquad \qquad q(t_F)
={1 \over 2}(Q_F + Q_F^\prime)\,.
\ee
hold.\footnote{Note that this is also the condition under which the classic action is actually 
minimal for $Q(t)=Q_c(t)$} In fact a crucial point is the integral (\ref{5.1.11}) in which the 
translational invariance of the measure $\mathcal D_c \xi(t)$ has been used. 

\quad In practice, once that eq. (\ref{5.1.14}) has been established, the milder assumption 
expressed by eq. (\ref{5.1.20}) is sufficient for what concerns $\xi(t)$. This enables us to avoid 
complicate discussions about the behaviour of $\tilde \xi (k)$ for $k\sim \omega$, to exploit the 
principal value prescription in (\ref{5.1.3}) and to include the border value $k=\omega$ and so 
the classical solutions explicitly in considerations.

\quad Now notice that, according to Eq. (\ref{2.2.22}), we have $\langle q(t)\rangle=
\langle \hat Q(t)\rangle _{\rm QM}$ and since $\mathcal L' (t)\, \hat Q(t)$ and 
$\mathcal L' (t) \,\hat P(t)$ vanish, as it can be immediately checked, we have, as in ordinary 
Quantum Mechanics,
\be
\lb {5.1.23}
 {d \over dt} \langle \hat Q(t) \rangle _{\rm QM} = \langle \hat P(t) \rangle _{\rm QM}  \,, 
\qquad \qquad  {d \over dt} \langle \hat P(t) \rangle _{\rm QM} 
= - \omega^2 \langle \hat Q(t) \rangle _{\rm QM}
\ee
So $ \langle q(t)\rangle$ is an exact solution of the classical equation of motion as expected,
 i.e.
\be
\lb{5.1.24}
\langle q(t)\rangle =\langle \hat Q(t)\rangle _{QM} = C \,\cos (\omega t + \delta) \,.
\ee 

\quad  Furthermore, if we introduce the time average (cf. eq. (\ref{3.1.18}))
\be
\lb{5.1.25}
      q_h(t)   =\int dt'  h(t-t' )  q(t')
\ee   		
we can write
\be
\lb{5.1.26}
 \langle q_h (t) \rangle = -i  {\partial \over \partial k} {\rm Tr}\left \{ \mathcal G(t_F,t_0;[kh])
\hat \rho_0\right \} |_{k=0}= \langle Q_h (t)\rangle _{\rm QM }
\ee
and, under the assumption of an effective support of $h(t)$ sufficiently small as in eq. 
(\ref{3.1.21}),
\bea
\lb{5.1.27}
&& \left \langle (q_h-\langle q_h \rangle)^2 \right \rangle = 
 -{\partial^2 \over \partial k^2} {\rm  Tr}\left \{\mathcal G( t_F,t_0;[kh]) \rho _0 \right \} 
|_{k=0}- \langle q_h \rangle ^2 = \nn \\
&&  \qquad \qquad \qquad ={G_{hh} \over \gamma} + 
\langle (q_h-\langle Q_h\rangle)^2 \rangle _{\rm QM}\,,
\eea
where
\be
\lb{5.1.28}
G_{hh}=\int dt \int dt'  h (t)\,G(t-t')\,  h (t')=\int dk \tilde h^*(k)\,
{\rm P}{1 \over k^2-\omega^2}\,\tilde h(k) \,,
\ee
$\tilde h(k)$ being the Fourier transform of $h(t)$.		

\quad More explicitly, according to eq. (\ref{2.2.27}) we can also write
\bea
\lb{5.1.29}
&& p(\bar q_1, h_1; \bar q _2, h_2;\dots \bar q_l, h_l )=\nn \\
&&= {1\over 2\pi}\,\int dk_1 dk_2 \dots dk_l \, 
e^{i(k_1 \bar q_1+ \dots k_l \bar q_l )} \, {\rm Tr}
\left \{ \mathcal G(t_F,t_0;[k_1 h_1+ \dots  k_l h_l ]) \hat \rho _0 \right \}= \nn \\
&& =\int dQ_F \int dQ_0 \,.  dQ_0'\, \langle Q_0,\, t_0 |\hat \rho_0 |Q_0',\, t_0 \rangle 
\int_{Q_0}^{Q_F} \mathcal D Q \int_{Q_0'}^{Q_F} \mathcal D Q' \, \exp \int_{t_0}^{t_F} dt \nn \\
&& \left \{ {-\gamma \over 4} \left [(\dot Q - \dot Q')^2 -
\omega^2 (Q-Q')^2 \right ]+ {i \over 2} (\dot Q^2 - \omega^2  Q^2 )
-{ i\over 2 }(\dot Q'^2 -\omega^2 Q'^2 )\right \} \nn \\
&& \left ({\gamma \over \pi}\right )^{1 \over 2} { 1 \over ({\rm det} G_{rs}))^{1/2}}
\exp \left \{-\gamma \sum_{rs}\left (\bar q_r - {Q_{h_r}+Q_{h_r}' \over 2}\right ) 
\right. \nn \\  
 && \left. \qquad \qquad \qquad \qquad \qquad\qquad \qquad G _{rs}^{-1}  
\left (\bar q_s -{Q_{h_s} +Q_{h_s}' \over 2} \right ) \right  \}\,  ,
\eea
with
\be
\lb{5.1.30}
G_{rs}= \int dt \int dt' \, h_r (t)\,G(t-t')\,  h_s (t').
\ee 

\quad Consistently with assumption (\ref{5.1.20}) $h(t)$ and $ h_1 (t), \dots h_l (t)$  must be such 
$G_{hh}$ or the matrix $G_{rs}$ be positive. Taking into account that
\be
\lb{5.1.31}
\int _{-\infty}^\infty dk\, {\rm  P} {1 \over k^2-\omega^2 } = 0\, ,
\ee 
it is clear that eq. (\ref{3.1.19})) does not longer provides correct choices. On the contrary an
admissible choice would be
\be
\lb{5.1.32} 
h(t)=A \, e^{-{t^2 \over \tau^2} } \cos \bar k t \,  ,
\ee
with $\tau \ll {1 \over \omega}$, $ \bar k >\omega$ and somewhat larger than 
${\sqrt 2 \over \tau}$\,. This is not a positive definite function but it is normalized 
to 1, if $A={1 \over \tau \sqrt \pi}\, e^{\bar k^2 \tau^2 \over 4}$ .
 
\quad Indeed, the Fourier transform of (\ref{5.1.32}) is
\be
\lb{5.1.33}
\tilde h(k)= {1 \over \sqrt {8\pi}} \, e^{\bar k^2\tau2 \over 4}\left (e^{-{\tau^2 \over 4} (k-\bar k)^2} + 
e^{ - {\tau^2 \over 4} (k+\bar k)^2}\right ) = {1 \over \sqrt {2\pi}}\,  e^{-{\tau ^2 k^2 \over 4}}\, 
\cosh {k \bar k \tau^2 \over2} \,.
\ee
For $ \bar k >{\sqrt 2 \over \tau} $  this quantity develops two symmetrical maxima which become 
soon very close to $-\bar k$ and $\bar k$ (for $\bar k={2 \over \tau}$ we have already 
$k_{\rm max} \simeq 0.95 \bar k$) and $G_{hh}$ becomes positive. So the factor 
$\cos \bar kt$ has simply the role of projecting $q(t)$ or equivalently $G(t-t')$ on a manifold in 
which the Fourier component with $|k|>\omega$ are dominant. Note that, if we introduce explicitly  
the corresponding restrictions on the spectrum of such functions, for $\bar k$ not too large 
(\ref{5.1.32}) is conceptually equivalent to a pure Gaussian.

\quad For the choice (\ref{5.1.32}) with $\omega \tau \ll 1$ and $\bar k \tau$ of the 
order of few units we find
\bea
\lb{5.1.34}
&& \langle q_h (t) \rangle = \int dt' h(t-t') \langle \hat Q(t')\rangle _{QM} =  \nn \\ 
&& \qquad = C\,e^{-{\omega^2 \tau2 \over 4}}\,  \cosh { \bar k \omega\tau^2 \over 2}\, 
\cos\, (\omega t + \alpha)\simeq C\,\cos\, (\omega t + \alpha)\,.
\eea
For what concerns $G_{hh}$ and so the intrinsic component of the fluctuation in (\ref{5.1.27}),
a rough estimate gives 
\be
G_{hh} \sim  {1 \over \bar k^2-\omega^2} \int dk\, \tilde h^*(k)\,\tilde h(k) =
{1 \over \tau \sqrt{8 \pi} }\, {e^{\bar k^2 \tau^2 \over 2}+1 \over \bar k^2-\omega^2}\,.
\ee 
This shows that for large frequency the spectrum of the fluctuations of the macroscopic
position $q(t)$ around the solution of the classical equation of motion 
diverges as the frequency increases; the average $q_h (t)$ over a time interval 
$\tau$  has the effect of damping such fluctuations. For $\bar k\sim 2/\tau$ we have simply
$G_{hh}\sim 1.7 /\tau\bar k^2 $. 

\quad Finally let us observe that, if in (\ref{5.1.29}, \ref{5.1.30}) $h_r (t)$ is identified with 
$\, h(t-t_r)\,$ with  $\,h(t)\,$ given by (\ref{5.1.32}) and $\, t_r-t_s \,$  large with respect 
to $\tau$ , the matrix results positive and nearly diagonal and eq. (\ref{2.1.3}) remain a 
good approximation.

% 5.2 5.2 5.2 5.2 5.2 5.2 5. 2 5.2 5.2 5.2

\subsection{Scalar Field} 

\quad We assume again (eq.(\ref {4.9}))
\be 
\mathcal L(x) \hat \rho = - {\gamma \over 4} \left ([\partial _\mu \hat \varphi (x),
[\partial ^\mu \hat \varphi (x), \hat \rho \,]] - m ^2 [\hat \varphi (x),
[\hat \varphi (x), \hat \rho \,]]\right )
\lb{5.2.1}
\ee
 and set
 \be
\mathcal K(x,j)\hat \rho = \mathcal L(x) \hat \rho - 
{i \over 2} j(x) \{ \hat \varphi (x), \hat \rho \, \} -  {1
  \over 2\gamma} \int _{\sigma_0} ^\sigma d^4 x^\prime \, G_m (x-
  x^\prime) \,j(x)\,j(x^\prime)\,, 
\lb{5.2.2}\int \mathcal D Q \int \mathcal D Q^\prime
\ee
where
\be
 G_m (x- x^\prime)={1 \over( 2 \pi)^4 } \int d^4 k \,
{\rm P}{1 \over k^2-m^2}\, e^{-i k_\mu (x-x^\prime)^\mu} 
\lb{5.2.3}
\ee
is the solution of the equation
\be
(\Box - m^2)\, G_m (x-x^\prime)=
\delta ^4(x-x^\prime)
\lb{5.2.4} 
\ee
and therefore the inverse of the differential operator $(\Box - m^2)$ under the appropriate 
restrictions.

\quad Then, it can be checked that the compatibility condition (\ref{3.2.2}) is still satisfied 
and all developments of the above subsection can be repeated with obvious modification, using 
the functional integrations as defined by eq. (\ref{3.2.12}). In this way, we arrive to the 
corresponding expression for the CFO, that directly in the continuous notation we can write
\bea
\lb{5.2.5}
&& \langle \varphi_F,\,\sigma_F | \mathcal G(\sigma_F,\sigma_0;[j])\,\hat \rho_0 
|\varphi'_F,\sigma_F \rangle = \nn \\
&& =\int \mathcal D_{\sigma_0}\varphi_0 \int \mathcal D_{\sigma _0} \varphi'\, 
\langle \varphi_0,\,\sigma_0|\hat \rho_0|\varphi'_0,\,\sigma_0\rangle 
\int _{\varphi_0 }^{\varphi_F} \mathcal D \varphi \int _{\varphi'_0}^{\varphi'_F} \mathcal D\varphi'\nn \\
&& \exp \int _{\sigma_0}^{\sigma_F} d^4 x \left \{ - {\gamma \over 4}
\left [\partial_\mu (\varphi-\varphi')\,\partial^\mu (\varphi-\varphi')- 
m^2 (\varphi-\varphi')^2 \right ]\right. \nn \\
&& -{i \over 2}\,  j (\varphi + \varphi')-{1 \over 4\gamma} \int_{\sigma_0}^{\sigma_F} d^4 x' 
 j(x) \,G_m (x-x')\,j(x') + \nn \\
&&\left. + {i \over 2} (\partial_\mu \varphi\, \partial^\mu \varphi - m^2 \varphi'^2 )
- {i \over 2} (\partial_\mu \varphi'\, \partial^\mu \varphi' - m^2 \varphi^2 )\right \} \, , 
\eea
where the functional integrals are defined as in (\ref{3.2.12}), but for convergence $\epsilon$ has 
to go to 0 faster than $\delta$. As a consequence, now, only the Fourier components of 
$\varphi (x)$, $\varphi'(x)$ and $j(x)$ with $ k^2\equiv (k_0 ^2 -{\bf k}^2) >m^2$ give 
contribution in the limit.In analogy with the preceding case, in practice,we may simply restrict 
the external source $j(x)$ by the condition
\be
\lb{5.2.6}
\int d^4x \, d^4 x' \,  j(x)\,  G_m (x-x' )\,  j(x') >0 \,  .
\ee
\quad From eq. (\ref{5.2.5}) we can derive the expression for 
${\bf f}(\sigma_F, \sigma_0; [\phi (x)])$ as a functional of the classic field $\phi(x)$ .
In the operator form we can write   
\bea
&& \qquad {\bf f}(\sigma_F, \sigma_0; [\phi (x)])\hat \rho  = C_\gamma\, 
\exp \left \{ - {\gamma \over 2}
\int d^4x  \right. \\
&&\left. [\partial _\mu (\phi\,(x)- \hat \varphi (x)) \,\partial
   ^\mu (\phi(x)- \hat \varphi (x)) -  
 m^2(\phi (x)-\hat \varphi (x))^2]\right \}\,\,\hat \rho \cdot \nn \\
&&  \exp \left \{ - {\gamma \over 2}
\int d^4x \,[\partial _\mu (\phi\,(x)- \hat \varphi (x))\, \partial
   ^\mu (\phi(x)- \hat \varphi (x)) -  
 m^2(\phi (x)-\hat \varphi (x))^2]\right \}. \nn
\lb{5.2.7}
\eea
\quad Eqs. (\ref{3.2.14}) and (\ref{3.2.18}) have to be replaced with
\be
\lb{5.2.8}
 \langle \phi_h (x) \rangle = \langle \hat \varphi_h (x)\rangle_{\rm QM}
\ee
and 
\be
\lb{5.2.9}
\left \langle \left ( \phi_h (x)- \langle \phi_h (x)\rangle \right)^2\right \rangle = 
{ 1 \over 2 \gamma}\, G_{hh}\,+\, \left \langle \left ( \hat\varphi_h (x)- 
\langle\hat \varphi_h (x)\rangle \right)^2\right \rangle_{\rm QM}\,, 
\ee
with
\be
\lb{5.2.10}
G_{hh}=\int d^4 x \,d^4 x' \,  h(x)\,  G_m (x-x' )\,  h(x' ) \,  ,
\ee                                                  
where it must be again  $G_{hh}>0$  and a permitted choice would be e.g.
\be
\lb{5.2.11}
h(x)= A\, e^{-{ t^2 \over \tau^2} - {{\bf x}^2 \over a^2}}\, 
\cos \bar k t \,  ,
\ee
now with  $\bar k^2 $  sufficiently larger than  $ m^2$ , $ \bar k > {\sqrt 2 \over \tau}$ and
$A={1 \over \pi^2 \tau a^3}\,e^{{1 \over 4}\bar k^2\,\tau^2}$ (note that in natural unity 
we may significantly assume $\tau \gg a$). With such choice $G_{hh} \sim {1 \over 4\pi^2\tau a^3}
{1 \over \bar k ^2-m^2} \, (e^{{1 \over 2}\bar k^2\,\tau^2} +1)$ and, if $\bar k \sim {2 \over \tau}$,
we have  $G_{hh}\sim 0.2\, {1 \over \tau a^3\bar k ^2}$.  
  
%6666666666666666666666666666666666666666666666666666666666666666666666666666666

\section{Electromagnetic field}

\quad If we want to introduce in a similar way a classical field in the case of the electromagnetic 
(e.m.) field, the form of $\mathcal L(x)$ is uniquely determined by Lorentz and gauge invariance, 
independently of a requirement of energy-momentum conservation. 

\qquad We must set
\bea
&\mathcal L(x)\hat\rho&= {\gamma \over 8}\, [\hat F_{\mu \nu}(x),\,[\hat F^{\mu
      \nu}(x),\hat\rho \,]]= \nn \\
&& = -{\gamma \over 4}\, \left ([\hat E^i (x),\,[\hat
    E^i(x),\hat\rho \,]] -[\hat B^i (x),\,[\hat B^i(x),\hat\rho \,]]\right )\,,
\lb{6.1}
\eea
where
 \bea
&& \quad F_{\mu \nu}(x) = \partial _\mu A_\nu (x) - \partial_\nu A_\mu (x)
\nn \\
&& E^i=F_{0i} \qquad \qquad B^i={1 \over 2}\epsilon _{ijl}\,F_{jl}\,.
\lb{6.2}
\eea

\qquad Eq. (\ref{6.1}) corresponds again to coefficients $\alpha_j$ of both signs in (\ref{1.2})
or (\ref{2.2.11}). So we are in a situation similar to the cases considered in sec. 5.

\qquad Actually we can set
\bea
\lb{6.3}
&&\mathcal K\left (x,\,j_\rho (x) \right)\hat \rho = \mathcal  L(x) \hat \rho + \\
&& \qquad \qquad + {i \over 2}\,j_\mu(x)\,
\{\hat A^\mu(x), \, \hat \rho \}+{1 \over 2 \gamma}\int_{\sigma_0}^\sigma d^4 x'\,j_\mu (x) \,
G^{\mu \nu}(x-x')\,j_\nu(x') \, \hat \rho \,, \nn
\eea
where $G^{\mu \nu}(x-x')$ is the Green function relative to the differential operator acting 
on the potential $A^\mu(x)$ in the equation of motion. This obviously depends on the 
gauge we use.

\qquad If we define $G_0(x-x')$ as in (\ref{5.2.3}) with $m=0$, in the Coulomb Gauge we have
\bea
\lb{6.4}
&& G^{00}(x-x')=-{1 \over 4 \pi}\,{1 \over |{\bf x} -{\bf x'}|} \, \delta(t-t')\,,
\qquad G^{0i}(x-x')=G^{i0}(x-x')=0 \,, \nn \\
&& G^{ij}(x-x')= - \left(\delta_{ij}-\partial_i \,{1 \over \nabla ^2}\,\partial_j \right ) 
G_0(x-x')= \nn \\
&&\qquad \qquad = -{1 \over (4 \pi)^4} \int d^4 k\, 
\left (\delta_{ij} -{k_hk_k \over {\bf k}^2}
\right ) {\rm P} {1 \over k^2}\,e^{-ik(x-x')}
\eea
and in a generic Lorentz gauge
\be
\lb{6.5}
G_\lambda^{\mu \nu}(x-x')={1 \over (4 \pi)^4} \int d^4 k\,\left (g^{\mu\nu} -(1-\lambda)\,{\rm P}
{k_\mu k_\nu \over k^2}
\right )\, {\rm P} {1 \over k^2}\,e^{-ik(x-x')}\,,
\ee
with the $\lambda$ specifying the specific gauge.

\qquad For consistency the classical source must be assumed to satisfy the continuity  equation
\be
\lb{6.6}
\partial_\mu j^\mu(x)=0
\ee
and, in analogy with the cases of the harmonic oscillator and of the scalar field, we can avoid 
an explicit reference to the complicate lattice formulation if we introduce the further 
restriction
\be
\lb{6.7}
\int d^4x\, d^4x' \,j_\mu(x) G^{\mu \nu}(x-x')j_\nu(x')=\int d^4 \, \tilde j_\mu^* (k)
\tilde G^{\mu \nu} (k) \tilde j_\nu (k) <0 \,.
\ee
\quad Under this assumption we can show that we can construct an operational density
${\bf f}(t_F,\,t_0;\,[f^{\mu \nu}])$, $ f^{\mu \nu}(x)=\partial^\mu a^\nu (x)- 
\partial^\nu a^\mu (x)$ being the classical e.m. field and $a^\mu(x)$ the classical 
tetra-potential, and from this to derive a positive probability distribution on the space of 
the histories of the classical field. On the other side we shall also see that eq. (\ref{6.1}) is consistent 
with local conservation of energy and momentum.

\quad These results hold both for the free e.m. field and for spinor Electrodynamics, what is 
much more interesting. Even i this second case  only the e.m. field is treated as continuously monitored 
and interpreted as a classical field, no reference is made to matter quantities. We shall treated separately 
the two cases.

%6.1 6.1 6.1 6.1 6.1 6.1 6.1 6.1 6.1 6.1 6.1 6.16.1 6.1 6.1  6.1 6.1 6.1 6.1 6.1 6.1

\subsection{Free field}

\quad The density of Lagrangian is
\be
L(x)=-{1 \over 4}\, F_{\mu \nu}(x)\, F^{\mu \nu}(x)={1 \over 2}\,({\bf E}^2(x) 
-{\bf B}^2(x))
\lb{6.1.1}
\ee
and  the energy momentum tensor
\be
\hat T_{\rm em}^{\mu \nu} (x) = \hat F^{\mu \rho} \hat
F_\rho^{\,\nu} + {1 \over 4} g^{\mu \nu}\hat F_{\rho \sigma}\hat
F^{\rho \sigma}\,.
\lb{6,1.2}
\ee
Specifically
\be
T_{\rm em}^{0 0} (x)= {1 \over 2}\,( {\bf E}^2+\hat {\bf
  B}^2)\,, \qquad \qquad T_{\rm em}^{0 i} (x)=( {\bf E}
\times \hat {\bf B})_i \,.
\lb{6.1.3}
\ee

\quad In the \textit{Coulomb gauge} we have the conditional equations 
\be
A_0(x)=0 \,, \qquad \qquad  \nabla \cdot {\bf A}(x)=0 \,, \qquad
\qquad \nabla\cdot {\bf E}(x)=0\,.
\lb{6.1.4}
\ee
and the \textit{Quantization rules}
\bea
&& [\hat A^i(t,{\bf x}),\,\hat A^j(t,{\bf x}')]=0\,, \qquad
  [\hat E^i(t,{\bf x}),\,\hat E^j(t,{\bf x}')]=0 \nn \\
&& [\hat E^i(t,{\bf x}),\,\hat A^j(t,{\bf x}')]
=i\,(\delta_{ij} -\partial _i \,{1 \over \nabla ^2}\,  \partial _j ) 
\delta ^3 ({\bf x} - {\bf x}')\, .
\lb{6.1.5}
\eea
From these we can derive the commutation relations for the field
\bea
&& [\hat E^i(t,{\bf x}),\,\hat E^j(t,{\bf x}')]=0\,, \qquad  
  [\hat B^i(t,{\bf x}),\,\hat B^j(t,{\bf x}']=0  \qquad \nn \\
&& [\hat E^i(t,{\bf x}),\,\hat B^j(t,{\bf x}')]
= \,-i\, \epsilon_{ijl} \partial_l\, \delta ^3 ({\bf x} -{\bf x}')\,,
\lb{6.1.6}
\eea
which are independent of the gauge.

\quad From eq, (\ref{6.1.6}) follows immediately that $\mathcal K\left (x,\,j_\rho (x) \right)$ 
as given by eqs. (\ref{6.1},\ref{6.3}) satisfies the consistency relation
\be
[\mathcal K\left (x,\,j_\rho (x) \right),\,\mathcal K\left (x',\,j_\rho (x') \right)]=0
\lb{6.1.7}
\ee
on a spacelike surface and also at equal time
\be
\lb{6.1.8}
\mathcal L^\prime(x')\, \hat T_{\rm em}^{0 \nu} (x) = 0 \, .
\ee
Consequently
\be
\partial _\mu \langle \hat T_{\rm em}^{\mu \nu} (x)\rangle _{QM}=0 \,.
\lb{6.1.9}
\ee

\quad According to standard methods for path integral in gauge field theories the CFO 
takes the form
\bea
\lb{6.1.10}
&&\langle {\bf A}_F, \sigma_F|\mathcal G (\sigma_F,\,\sigma_0;\,[j_\rho])
|{\bf A}'_F,\, \sigma_F \rangle ={ \rm const} \int \mathcal D_{\sigma_0} {\bf A}_0  \\
&& \int \mathcal D_{\sigma_0} {\bf A}'_0 \,
\langle {\bf A}_0,\, \sigma_0 | \hat \rho_0 |{\bf A}'_0,\, \sigma_0 \rangle
\int_{{\bf A}_0}^{{\bf A}_F} \mathcal D {\bf A}\, \,\delta[\nabla \cdot {\bf A}]
\int_{{\bf A}'_0}^{{\bf A}'_F} \mathcal D{\bf A}'\,\, \delta[\nabla \cdot {\bf A}'] \nn \\
&& \exp \int _{\sigma_0}^{\sigma_F} d^4 x \left \{ {\gamma \over 16 }(F_{\mu \nu}- F'_{\mu \nu})
(F^{\mu \nu}- F'^{\mu \nu})) 
+ {i \over 4} ( F_{\mu \nu}F^{\mu \nu}- F'_{\mu \nu}F'^{\mu \nu}) - \right.\nn \\
&& \left. - {i\over 2}\,j_\mu \,(A^\mu + A'^\mu)+ {1 \over 2\gamma} 
\int _{\sigma_0}^{\sigma_F} d^4 x' \, j_\mu(x)\, G^{\mu \nu}(x-x')\, j_\nu (x') \right \}\,,\nn
\eea
where consistently with (\ref{6.6}) we can assume
\be
\lb{6.1.11}
j_0(x)=0 \,, \qquad \qquad \nabla \cdot {\bf J}(x) = 0
\ee
and then (\ref{6.4}) becomes 
\be
\lb{6.1. 12}
\int d^4 k \, \tilde j^*(k) \left ( \delta_{ij} - {k_i k_j \over {\bf k}^2} \right )j_j(k)\,
{\rm P}\, {1 \over k^2} = \int d^4 k \,|{\bf j}(k)|^2 {\rm P}\, {1 \over k^2}>0 \,,
\ee
which essentially implies that timelike $k$ prevail on spacelike ones. 

\quad From (\ref{6.1.10}) we obtain by the usual procedure 
\bea
\lb{6.1.13}
&& {\bf f}(\sigma_F,\,\sigma_0;\,[a_\rho(x)])\,\hat \rho =\\
&& = \int \mathcal D_c {\bf j}\, \delta[\nabla \cdot {\bf j}]\, 
  \exp \left (i \int d^4x \,j_\mu (x)\, a^\mu (x)\right )
\mathcal G (\sigma_F,\,\sigma_0;\,[j])\hat \rho = \nn \\
&& = C_\gamma \, \exp \left \{ {\gamma \over
  4}\int d^4x \,[f _{\mu \nu}(x)- \hat F_{\mu \nu}(x)][f ^{\mu \nu}(x)-
      \hat F^{\mu \nu}(x)] \right \}\, \hat \rho \nn \\
 && \qquad \qquad  \qquad \exp \left \{ {\gamma \over
  4}\int d^4x \,[f _{\mu \nu}(x)- \hat F_{\mu \nu}(x)][f ^{\mu \nu}(x)-
      \hat F^{\mu \nu}(x)] \right \} \,, \nn
\eea
which is obviously positive.

% 6.2 6.2 6.2 6.2 6.2 6.2 6.2 6.2 6.2 6.2 6.2 6.2 6.2 6.2 6.2

\subsection{Spinor Electrodynamics}

\quad Now let us consider the more significant model of spinor electrodynamics and assume 
as classical variables the e. m. field components alone as in the free case. 

\quad Any number of Dirac fields could be included in principle, however for
 simplicity we shall  explicitly write only one field. We stress that we do not 
consider any classical variable relative to the Dirac field and any observation 
on the system is supposed to be expressed in terms of modifications on the 
classical e. m. field (see App. D an explicit discussion).

\quad The Lagrangian density of the system is
\bea
\lb{6.2.1}
&& L(x)=-{1 \over 4}\, F_{\mu \nu}\, F^{\mu \nu}-{i \over 2} (\bar \psi \gamma^\rho 
\partial_\rho \psi- \partial_\rho \bar \psi \gamma^\rho \psi)- \nn \\
&& \qquad \qquad \qquad \qquad 
- m\, \bar \psi \psi -eA^\mu \, \bar \psi \gamma ^\mu \psi \,.
\eea

\quad We shall use the same convention as in the preceding subsection and shall 
operate again in the Coulomb gauge. Then instead of (\ref{6.1.4}) we have
\be
\lb{6.2.2}
\nabla \cdot {\bf A}(x)=0 \,, \qquad\qquad {\rm but } \qquad \qquad 
\nabla\cdot {\bf E}(x)=e \bar \psi \gamma^0 \psi
\ee
and $ {\bf E}(x)= {\bf E}_{\rm T}(x)+{\bf E}_{\rm L}(x)$ with
\bea
\lb{6.2.3}
&&\,\,\,{\bf E}_{\rm T}(x)=-\, {\partial {\bf A}(x) \over \partial t}\,, \qquad 
{\bf E}_{\rm L}(x)= -\,\nabla A^0(x)\,, \nn \\
&& A^0 (t, \, {\bf x})=-\, {e \over 4 \pi}\int d^3 {\bf y}\,{1 \over |{\bf x}-{\bf y}|}\,
\bar \psi (t, \, {\bf y}) \gamma ^0 \psi (t, \, {\bf y}) \,. 
\eea

\quad The basic commutation rules for the e. m. field are again given by eqs. 
(\ref{6.1.5},\ref{6.1.6}) with the electric field replaced by its transverse part
$E_h^{\bf T}(x)$. The remaining commutation rules are obviously
\bea
\lb{6.2.4}
&& \{\hat \psi_\alpha(t,\,{\bf x}), \,\hat \psi_\beta(t,\,{\bf x}') \}=
\{\bar{\hat\psi}_\alpha(t,\,{\bf x}), \,\bar{\hat\psi}_\beta(t,\,{\bf x}') \}=0  \nn \\
&& \{\hat\psi_\alpha(t,\,{\bf x}), \,\bar{\hat \psi}_\beta(t,\,{\bf x}') \}=\gamma_{\alpha \beta}^0
\delta^3({\bf x}-{\bf x}')\,,\nn \\
&& [ \hat \psi_\alpha(t,\,{\bf x}), \,\hat A^i(t,\,{\bf x}')]=
[\bar {\hat \psi}_\alpha(t,\,{\bf x}), \,\hat A^i(t,\,{\bf x}')] =0 
\eea
and so
\be
\lb{6.2.5}
[ \hat \psi_\alpha(t,\,{\bf x}), \,\hat E^i_{\bf T}(t,\,{\bf x}')]=
[\bar{\hat \psi}_\alpha(t,\,{\bf x}), \,,\hat E^i_{\bf T}(t,\,{\bf x}')] =0\,.
\ee
\quad From the preceding equations there follow immediately
\bea
\lb{6.2.6} 
&& [ \hat \psi_\alpha(t,\,{\bf x}), \,\hat A^0(t,\,{\bf x}')]=-\,{e \over 4 \pi}\, 
{1 \over |{\bf x}-{\bf x}'|} \,\psi_\alpha(t,\,{\bf x}) \nn \\ 
&& [ \bar{\hat \psi}_\alpha(t,\,{\bf x}), \,\hat A^0(t,\,{\bf x}')]=\,\,\, 
{e \over 4 \pi}\,{1 \over |{\bf x}-{\bf x}'|} \, \bar{\hat \psi}_\alpha(t,\,{\bf x})
\eea
and
\be
\lb{6.2.7} 
[ \hat A^0(t,\,{\bf x}), \,\hat A^i(t,\,{\bf x}')]=0 \,, \qquad \qquad 
[\hat A^0(t,\,{\bf x}), \,\hat E_{\rm T}^i(t,\,{\bf x}')]=0\, .
\ee
Then, (\ref{6.1.6}) holds also for the total e. m. field, as should be by gauge invariance, 
$A^0(t,\,{\bf x})$ and consequently the longitudinal and the total electric fields 
commute with a bilinear expression of the form
$\bar {\hat \psi}_\alpha(t,\,{\bf x}')\hat \psi_\beta(t,\,{\bf x}')$
\be 
\lb{6.2.8} 
[\hat E^i(t,\,{\bf x}), \, \bar {\hat \psi}_\alpha(t,\,{\bf x}') 
\hat\psi_\beta(t,\,{\bf x}')]=0\,.
\ee
\quad Since only the e. m. field is introduced as a classic variable  $\mathcal L (x)$ 
and $\mathcal K(x,j)$ are again as defined by eqs. (\ref{6.1},\, \ref{6.3}) and
the restriction on $j^\mu(x)$ is again expressed by eq. (\ref{6.7}). The CFO and 
the density of operation remain of the form (\ref{6.1.10}) and  (\ref{6.1.13}) but 
with the free e. m. Lagrangian replaced by the complete Lagrangian (\ref{6.2.1}), 
the functional integral understood even on Clifford field $\psi$ and $\bar \psi$, 
evolution of the operators in the Heisenberg picture intended with respect to the 
total Hamiltonian $\hat H=\hat H_{\rm em}+\hat H_{\rm D}+\hat H_{\rm int}$.

\quad Correspondingly the energy momentum tensor can be written
\be
\lb{6.2.9}
  T^{\mu \nu}(x)= T_{\rm em}^{\mu \nu}(x)+ T_{\rm D}^{\mu \nu}(x)+
 T_{\rm int}^{\mu \nu}(x) \,,
\ee
where $ T_{\rm em}^{\mu \nu}(x)$ is again of the form (\ref{6.1.1}),
\bea
\lb{6.2.10}
&&  T_{\rm D}^{\mu \nu}(x)= {i \over 2}(\bar \psi \gamma^\mu \partial ^\nu \psi - 
\partial^\nu \bar \psi \gamma^\mu \psi)- \nn \\
&& \qquad \qquad \qquad -g^{\mu \nu} [{i \over 2}(\bar \psi \gamma^\rho \partial_\rho \psi -
\partial_\rho \bar \psi \gamma^\rho \psi) -m\bar \psi \psi]
\eea
and
\be
\lb{6.2.11}
 T_{\rm int}^{\mu \nu}(x)= g^{\mu \nu} e \bar \psi \gamma^\rho \psi A_\rho\,.
\ee
\quad Then from the commutation rules (\ref{6.1.5}, \ref{6.1.6}, \ref{6.2.4}-\ref{6.2.8}) 
one can immediately check that again 
\be
\lb{6.2.12}
\mathcal L^\prime(x')\, \hat T^{0 \nu} (x) = 0 \, .
\ee
and so the local energy momentum conservation remains valid in the form
\be
\partial_\mu \langle \hat T^{\mu \nu} (x)\rangle _{QM}=0 \,.
\lb{6.2.13}
\ee 
also in this case. Similar conservation equation obviously are valid for 
the electric charge, the barionic number, etc.

\quad Notice that the counterpart of (\ref{6.1.10}) can be immediately rewritten in a generic
Lorentz gauge as
\bea
\lb{6.2.14}
&&\langle {\bf A}_F,\,\zeta_F,\, \sigma_F|\mathcal G (\sigma_F,\,\sigma_0;\,[j_\rho])
|{\bf A}'_F,\, \zeta'_F, \,\sigma_F \rangle = \nn \\
&&\qquad  =\sum_{\zeta_0 \zeta'_0} \int \mathcal D_{\sigma_0} {\bf A}_0 \,\mathcal D_{\sigma_0} {\bf A}'_0
\langle {\bf A}_0,\, \zeta_0, \,\sigma_0 | \hat \rho_0 |{\bf A}'_0,\, \zeta'_0, \,\sigma_0 \rangle
\nn \\
&& \qquad \int_{{\bf A}_0}^{{\bf A}_F} \mathcal D A\, \,\mathcal D \bar \psi \, \mathcal D \psi
\, \int_{{\bf A}'_0}^{{\bf A}'_F} \mathcal D A'\,\,\mathcal D \bar \psi' \, \mathcal D \psi'  \nn \\
&& \exp \int _{\sigma_0}^{\sigma_F} d^4 x \left \{ {\gamma \over 8 }(F_{\mu \nu}- F'_{\mu \nu})
(F^{\mu \nu}- F'^{\mu \nu}))  - \right.\nn \\
&& \left. - {i\over 2}\,j_\mu \,(A^\mu + A'^\mu)+ {1 \over 4\gamma} 
\int _{\sigma_0}^{\sigma_F} d^4 x' \, j_\mu(x)\, G^{\mu \nu}(x-x')\, j_\nu (x') + \right . \nn  \\ 
&& \qquad \qquad \left. +i[L_{\rm eff}(A,\,\bar \psi, \, \psi)-
L_{\rm eff}(A',\,\bar \psi', \, \psi')]\right \}\,,
\eea
where $\zeta_0,\, \zeta'_0,\,\zeta_F,\,\zeta'_F$ specify initial and final 
states of the spinor field,
\be
\lb{6.2.15}
L_{\rm eff}(A,\,\bar \psi, \, \psi)=L(A,\,\bar \psi, \, \psi)-{1 \over 2\lambda}(\partial_\mu A^\mu)
\ee
and $G_\lambda^{\mu \nu} (x-x')$ defined by eq. (\ref{6.5}).

\quad Finally let us consider the fluctuations of the classical e. m. field around its expectation
value. Formally we can write
\bea
\lb{6.2.16}
&&(-i)^2 {\delta \over \delta j^\mu(x)}\, {\delta \over \delta j^\nu(x')} \, {\rm Tr}\left.
\left [\mathcal G (\sigma_F,\,\sigma_0;\,[j_\rho])\hat \rho_0\right ] \right |_{j=0} = \nn \\
&&= {1 \over 2 \gamma} \, G_{\mu \nu}(x-x') + \theta (t-t') \,{1 \over 2}\, {\rm Tr}
\left [ \hat A_\mu (x) \mathcal G (\sigma,\,\sigma')\,
\{\hat A_\nu (x'),\, \mathcal G (\sigma',\,\sigma_0) \hat \rho_0\} \right ] +\nn \\
&& \qquad + \theta (t'-t) \,{1 \over 2}\, {\rm Tr}
\left [ \hat A_\nu (x') \mathcal G (\sigma',\,\sigma)\,
\{\hat A_\mu (x),\, \mathcal G (\sigma,\,\sigma_0) \hat \rho_0 \} \right  ] \cong \nn \\
&& \qquad \cong  {1 \over 2 \gamma} \, G_{\mu \nu}(x-x')\,+\, \langle A_\mu(x)\, 
A_\nu (x')\rangle _{\rm QM}\, ,
\eea
if $t'$ is sufficiently close to $t$, and
\bea
\lb{6.2.17}
&&\langle f_{\mu \nu}(x)\, f_{\rho \sigma}(x')\rangle ={1 \over 2 \gamma }
\left [\partial_\mu \partial_\rho '\,G_{\nu \sigma}(x-x')-
\partial_\mu \partial_\sigma ' \, G_{\nu \rho} (x-x') - \right. \nn \\
&&\qquad \left.  -\partial_\nu \, \partial _\rho G_{\mu \sigma} (x-x')+ 
\partial _\nu \, \partial_ \sigma \,  G_{\mu \rho}(x-x')(x-x')\right ]+ \nn \\
&& \qquad +\, \langle \hat F_{\mu \nu}(x)\, \hat F_{\rho \sigma}(x')\rangle _{\rm QM}\,.
\eea
\quad Then, let us set  $ E_{\rm class} ^i (x)=f_{0i}(x)$,
$B_{\rm class} ^i (x)={1 \over 2} \epsilon _{ijl}\,f_{jl}(x)$ and 
\be
\lb{6.2.18}
{\bf E}_h (x)=\int d^4x'\, h(x-x')\,{\bf E}_{\rm class}(x')\, \qquad 
{\bf B}_h (x)=\int d^4x'\, h(x-x')\,{\bf B}_{\rm class}(x') \,,
\ee
with $h(x-x')$ as in (\ref{5.2.11}). 
Obviously we have
\be
\lb{6.2.19}
\langle {\bf E}_{\rm class}(x) \rangle = \langle \hat {\bf E}(x) \rangle_{\rm QM}\, ,
\qquad \langle {\bf B}_{\rm class}(x) \rangle = \langle \hat {\bf B}(x) \rangle_{\rm QM}\,
\ee 
and, again for $\bar k \sim 2/\tau$,   
\bea
\lb{6.2.20}
&& \langle (E_h^i - \langle  E_h^i \rangle )^2\rangle ={1 \over 2\gamma}\int
d^4k \,{\rm P}\,{k_0^2-k_i^2 \over k^2}\,|\tilde h(k)|^2  \,+\, 
\langle (\hat E_h^i - \langle  \hat E_h^i \rangle )^2\rangle_{\rm QM}\sim \nn \\
&& \qquad \sim {0.2 \over \gamma\, \tau \,a^3} +\langle (\hat E_h^i - \langle \hat  E_h^i 
\rangle )^2\rangle_{\rm QM}\,,
\eea
A similar equation can be derived for $B_h^i$.

%7777777777777777777777777777777777777777777777777777777777777777777

\section {Conclusive consideration}

\quad In conclusion we have shown on three different models that it is possible modify
the formalism for the continuous monitoring of {\it macroscopic quantities} in 
Quantum Theory in such a way that the basic conservation laws are preserved.

\quad As we mentioned the idea is that Quantum Theory should be modified by introducing certain  
basic macroscopic quantities, that are formally treated as continuously monitored, but are 
actually thought as {\it classical quantities} or {\it beables}. These are supposed to have well 
determined values at each time and in terms of modifications of them any other observation 
should be expressed.

\quad Obviously  the most significant of the models we have proposed is the 
spinor electrodynamics, in which the macroscopic electromagnetic field components are considered as 
{\it classical}. Even this to be made realistic should be extended at least to the so called 
particle {\it Standard Model}. In the usual formulation of the latter, in which the Higgs 
is treated as elementary, this may be not a trivial task. The difficulty comes from 
the occurrence of terms quadratic in the e.m. potential in the boson sector of the theory. 
However there are some important properties of the model that should remain valid in a more 
complete theory; let us briefly discuss them.

\quad First of all, note that it is implicitly built in the eqs. (\ref{2.1.3}-\ref{2.1.7}) (that remain 
valid at least as good approximations) that, when applied to a small number of particles, the 
formalism reproduces the usual quantum theory. Two essential modifications occur:

\quad 1) only observables that can be expressed in a modification of the above macroscopic field 
have to be considered;

\quad 2) the usual unitary evolution has to be corrected by the action of the mapping 
$\mathcal G (t,\,t_a)$ on the initial density operator $\hat \rho(t_a)$.

In the context, small number of particles means compatible with a negligible macroscopic 
e. m. field.

\quad A detailed discussion of the question is given in App. D. Here we want rather comment 
the meaning of the two statements.

\quad Point 1). This should raise no problems. In fact, practically all our 
particle detectors work in terms of e. m. effects, that by appropriate amplification reach 
the macroscopic scale. In last analysis, even in the spirit of von Neumann  
psycho-physical parallelism, the states of our brain related to our perceptions are 
expressed in terms of membrane potentials, action potentials, charge distributions and so on.

\quad Point 2). The entity of the corrections to time evolution is controlled by the value of the 
constant $\gamma$, that has to be intended as a new fundamental constant of nature It is clear 
that, since  ordinary quantum theory works well for few particles, $\gamma$ should be small. 
On the other side in eq. (\ref{6.2.19}) $\gamma$ occurs in the denominator of the variance of 
the field. Now it is clear that, in order the all idea to make sense, such variance has 
to be negligible at some typical macroscopic scale. That is for some reasonable values of $\tau$ 
and $a$ in (\ref{6.2.19}) and some
appropriate ${\bf E}_{\rm typ}$ or ${\bf B}_{\rm typ}$  we must have
\be
\lb{7.1}
\langle ({\bf E}_h-\langle {\bf E}_h\rangle)^2\rangle /  {\bf E}_{\rm typ}^2 \ll 1\, , \qquad
\langle ({\bf B}_h-\langle {\bf B}_h\rangle)^2\rangle /  {\bf B}_{\rm typ}^2 \ll 1 \,.
\ee
This provide us the lower bound
\be 
\lb{7.2}
\gamma \gg 0.2/ \left ({\bf E}_{\rm typ}^2\,\tau \,a^3 \right )  \,.
\ee
and a similar for ${\bf B}_{\rm typ}$. To see what this means let us take 
${\bf E}_{\rm typ}^2\sim {\bf B}_{\rm typ}^2$ as the 
value of equilibrium inside a cavity at ordinary temperature $T=300\,{\rm K}$ and e. g.   
$\tau = 1\,{\rm ms} = 3 \times 10^7\,{\rm cm} $, $a=1\,{\rm \mu m} = 10^{-4}\,{\rm cm}$. 
For the density e. m. energy the Stefan-Boltzman law gives in natural units
\be
\lb{7.3}
u(T)=7.56 \times 10^{-15} \, T^4 \,{\rm erg \,\, cm^{-1}}\, =2.39 \times 10^2 
\,T^4 \,{\rm cm^{-4}}\,.
\ee
Then, setting
\be
\lb{7.4}
{\bf E}_{\rm typ}^2 \sim {\bf B}_{\rm typ}^2 \sim u(300\,\,{\rm K})=1.44 \times 10^{18}\,{cm}^{-4}\,,
\ee
 eq. (\ref{7.2}) becomes
\be
\lb{7.5}
\gamma \gg 5\times 10^{-15}\,
\ee
which should not raises problems too.

\quad Second, to make a comparison with the collapse models, let us observe that from the mathematical 
point of view our proposal corresponds to a specific choice of the ``dissipative'' term in the 
Liouville-von Newman equation, dropping the positivity requirement at the price of restricting 
the class of the observables.The result is the possibility to reformulate the theory in such a way 
that the interference terms among macroscopic states do not decay but are conceptually suppressed.    

\quad Finally let us stress that an equation of the type (\ref{1.3}) breaks temporal inversion 
invariance and this should have astrophysical and cosmological consequences. Possibly it is just by 
such a kind of consequences that a theory of this type could be tested and the new fundamental constant 
$\gamma$ determined. More in general, for what concerns experimental tests see e. g. the discussion in 
\cite{collapse} for the case of collapse models, that in part should apply even to the present one. 

%rrrrrrrrrrrrrrrrrrrrrrrrrrrrrrrrrrrrrrrrrrrrrrrrrrrrrrrrrrrrrrrrrrrr    

\section{Acknowledgements}
\quad Warm thanks are due to my friend L. Lanz for many interesting discussion and critical remarks.   

%aaaaaaaaaaaaaaaaaaaaaaaaaaaaaaaaaaaaaaaaaaaaaaaaaaaaaaaaaa

\section{Appendices}

\appendix 

\section {Path integral for the harmonic oscillator amplitude}

\quad Reintroducing the mass explicitly the Lagrangian of the harmonic oscillator 
can be written
\be
\lb{A.1}
 L={m \over 2} (\dot Q ^2 - \omega^2 Q^2)
\ee
and we have for the ordinary transition amplitude for an infinitesimal time integral 
\be
\lb{A.2}
   \langle Q',\,t+\epsilon|Q, \, t \rangle = 
\sqrt  {m \over 2 \pi i \epsilon} \exp \left \{i\, {m \over 2} 
\left [{1 \over \epsilon}\, (Q'-Q)^2 - \epsilon \, \omega^2 \, Q^2 \right ] \right \}\,. 
\ee
Then, in our notation the path integral expression for a finite time interval is
\bea
\lb{A.3}
&  \langle Q,\, t |Q_0, \, t_0 \rangle = \left ({m \over 2 \pi i \epsilon}\right )^{N/2}
 \int dQ_1\,dQ_2 \, \dots\, dQ_{N-! } \qquad \nn \\ 
&\qquad \qquad \exp \left \{i\, {m \over 2} 
\sum _{j=0}^{N-1}\left [{1 \over \epsilon}\, (Q_{j+1}-Q_j)^2 - 
\epsilon \, \omega^2 \, Q_j^2 \right ] \right \}\,,
\eea
the limit for large $N$ being obviously understood. 

\quad The above integral can be explicitly performed step by step and we obtain
\be
\lb{A.4}
\langle Q,\, t |Q_0, \, t_0 \rangle = \sqrt {m \omega \over 2 \pi i \sin \omega \tau}\,
\exp \left \{im \,\omega\, {(Q^2+Q_0^2)\,\cos \omega \tau - 2QQ_0 \over 
2 \sin \omega \tau}\right \}\,,
\ee
where $\tau=t-t_0$ \cite{FH}. 

\quad To prove this is sufficient to observe that (\ref{A.4}) reduces to (\ref{A.2}) for an 
infinitesimal interval and it reproduce itself after a further infinitesimal step
\bea
\lb{A.5}
&\langle Q',\,t+\epsilon|Q_0, \, t_0 \rangle = \int dQ \,
\langle Q',\,t+\epsilon|Q, \, t \rangle \langle Q,\,t|Q_0 \, t_0 \rangle = \nn\\
&={m \over 2 \pi i} \sqrt{\omega \over\epsilon \sin \omega \tau} \int dQ\, \exp \left \{i{m \over 2} 
\left [\left({1 \over \epsilon} -\epsilon \omega^2 +
\omega {\cos \omega \tau \over \sin \omega \tau}\right)\,Q^2  \right. \right. \qquad 
\qquad \nn \\
& \qquad \qquad \left. \left. - 2 \left ({1 \over \epsilon} Q' + 
{\omega \over \sin \omega \tau} \, Q_0 \right)\,Q +i{1 \over \epsilon} Q'^2 +
\omega {\cos \omega \tau \over  \sin \omega \tau} \, Q_0^2 \right ] \right \}= \nn \\
& =\sqrt {m \omega \over 2 \pi i\, \sin \omega(t+\epsilon)} \, \exp \left \{ im \omega \, 
{(Q'^2+Q_0^2)\,\cos \omega (\tau+\epsilon) \, - 2Q'Q_0 \over 
2 \sin \omega (\tau+ \epsilon)}\right \}\,,
\eea
up to terms of order $\epsilon^2$ in the last equality.

\quad For real $m$, actually the the above integral would be undetermined and it is 
usually made well defined by performing an infinitesimal rotation in the $t$ complex plane.  
On the contrary, note that, due to the prevalence of the term ${1\over \epsilon}$ in the
 coefficient of  $Q^2$ everything becomes perfectly defined if we put $m=i \mu$ with $\mu>0$,
independently of the sign of the other two terms. Then the situation becomes strictly 
similar to the one encountered in eqs (\ref{3.1.9}) and (\ref{3.1.10}).

\section {Positivity of the basic matrix}

\quad Let us denote by $K_{ij}^{(n)}(\lambda)$ the $n \times n$ matrix
\be
\lb{B.1} 
K_{i\,j}^{(n)}(\lambda)
=\left(\matrix {1-\lambda & -1 & 0 &\dots & 0 & 0\cr
 -1 &  2-\lambda & -1 & \dots & 0  & 0 \cr
 0 & -1 & 2-\lambda & \dots & 0 & 0\cr
\dots & \dots & \dots & \dots & \dots &\dots \cr
0 & 0 &  0 & \dots & 2-\lambda &-1\cr
0 & & 0 &0 \dots & -1 & 1 \cr} 
 \right )\,,
\ee
by $ \bar K_{ij}^{(n)}(\lambda)$ the $n\times n$ obtained by suppressing the last row 
and column in $K_{ij}^{(n+1)}(\lambda)$ and by $\bar{\bar K}_{ij}^{(n)}(\lambda)$ the 
matrix obtained by suppressing the first row and column in $\bar K_{ij}^{(n+1)}(\lambda)$.
 Obviously, with reference to the main text, we have $\bar K_{ij}=K_{ij}^{(N)}
(\epsilon^2 \omega ^2)$.

\quad Let us further denote by $D_n$, $\bar D_n$, $\bar {\bar D}_n$ the determinants of 
$ K_{ij}^{(n)}(0)$, $ \bar K_{ij}^{(n)}(0)$ and $ \bar {\bar K}_{ij}^{(n)}(0)$  respectively.
By developing them  with respect to the first row, we obtain the recurrence 
relations
\be
\lb{B.2}
D_n = \bar D_{n-1}-\bar D_{n-2}\,, \quad
\bar D_n = \bar{\bar D}_{n-1}-\bar {\bar D}_{n-2}\, , \quad 
\bar {\bar D}_n = 2\, \bar{\bar D}_{n-1}-\bar {\bar D}_{n-2}\, .
\ee
By induction from the last equation we have
\be
\bar {\bar D}_n=n+1
\ee
and by using such result from the second and the first equation
\be
\bar D_n =1 \quad {\rm and}\quad D_n=0\,.
\ee. 
\quad Let us now consider the matrix  $\bar K_{ij}^{(n)}(0)$ and note that this is
positive, since all its principal subdeterminants are of the type $\bar D_p$ or 
$\bar {\bar D}_p$ with $p \leq n$ and so their are positive (we call principal 
determinants those obtained suppressing any number of rows and the corresponding 
columns in the original determinant). Then the eigenvalues of $\bar K_{ij}^{(n)}(0)$ are 
all positive and are given by the roots of the polynomial 
\be
\lb{B.3}
 \det \bar K_{ij}^{(n)}(\lambda)\equiv \det ( \bar K_{ij}^{(n)}(0) -\lambda \delta_{ij})
= A_0 - A_1 \lambda + A_2 \lambda^2- \dots +(-\lambda)^{n-1}\,,
\ee
where the coefficients $A_0,\, A_1, \, \dots $ are expressed as sum of principal 
determinant and are again positive.  In particular we have
\be
\lb{B.4}
A_0 =\bar D_n =1, \qquad A_1= \bar {\bar D}_{n-1}+(n-2)\bar D_{n-2} =2(n-1).
\ee
Notice that, in order the matrix $\bar K_{ij}^{(n)}(\lambda)$ be itself positive, 
$\lambda$ must be smaller than the minimum  eigenvalue $\lambda_{\rm m}$ of 
$\bar K_{ij}^{(n)}(0)$ and, since $A_2>0$, it must be $\lambda_m>{A_0 / A_1} = 1/2(n-1)$.
Then, if 
\be
\lb{B.5}
\omega^2 \,\epsilon^2 \equiv {\omega ^2\, T^2 \over N^2}<{1 \over 2( N-1)} \quad 
{\rm for} \quad N>2\, \omega^2 \,T^2 \,,
\ee
$\bar K_{ij}\equiv \bar K_{ij}^{(N)}(\epsilon^2 \omega^2) $ is certainly positive, 
as we stated.

\section{Fourier transform of the interpolating continuous world line}

\quad Let us consider the continuous function defined in the entire interval $(t_0,\, t_F)$ by
\be
\lb{C.1}
Q(t)=Q_{j-1}+(t-t_{j-1}){Q_j-Q{j_1} \over \epsilon}\, \qquad {\rm for}\quad t \in (t_{j-1},\,t_j)\,,
\ee
for every $j=0,\,1,\,...\, N$. This function interpolates the values $Q_j$ and makes the exponential 
in eq.(\ref{3.1.12}) identical to those in eq.(\ref{3.1.10}) up to to a terms of order of 
$\epsilon$ coming from the potential part of the action.

\quad The Fourier coefficients of such function are given by
\bea
\lb {C.2}
&\tilde Q _k = {1 \over \sqrt T} \int _{t_0}^{t_F} dt \, e^{-ikt}\, Q(t) = 
\sum_{j=0}^{N-1} e^{-ik{t_{j+1}+t_j \over 2}} \qquad \qquad \qquad \\
& \qquad \qquad \left [ {Q_{j+1}+ Q_j \over 2}\,\epsilon\,{\sin k\epsilon / 2 \over k\epsilon/2} + 
(Q_{j+1}-Q_j)\,{i \over k}
\left(\cos k\epsilon /2 -{\sin k\epsilon/2 \over k\epsilon /2} \right  ) \right ] \nn \,.
\eea
The first term in the above expression is important for small $k$ but it is of the order 
of $\epsilon$, the second term is of the order of unity and it is important for 
$k\epsilon \sim \pi$ and so for $|k|\gg \omega$ if $\epsilon \ll {2\pi \over \omega}$.
Consequently the region $|k|<\omega$ gives a vanishing contribution for $\epsilon \to 0$ 
($N\to \infty$).

\section{Ricovery of ordinary Quantum Mechanics for a small system}

\quad In the perspective of the paper any observation on a system has to be expressed in terms 
of the modification that the system induces on the {\it classical} e. m. field.
 
\quad Let us  consider,  
e. g., a system of a small number of particles characterized by a certain set of invariants 
(a total electric charge, baryon number, lepton number, etc) to which we shall refer as the 
object system. Let us assume that such particles interact freely among themselves 
during a certain interval of time $(t_a,\,t_b)$. We admit any kind of rearrangement inside 
the system, exchange of energy and momentum, production or destruction of particles, but no 
interaction with the external environment during such interval of time.

\quad We assume that at the time $t_b$ the system comes in contact with an apparatus, by which 
the specific type of final particles, their momenta, energies etc. can be detected. To be specific 
we may think of the apparatus as a set of counters, filling densely a certain region kept under 
the action of a magnetic field.

\quad Both the object system and the apparatus in their specific states must be thought as states 
of the same system of fields and can be expressed as appropriate composed creator operators applied
to the vacuum state. Let us denote by $|u_1(t)\rangle ;\,|u_2(t)\rangle,\, \dots$ and 
$|U_1(t)\rangle ;\,|U_2(t)\rangle,\, \dots$ two orthogonal basis in the subspaces of the object 
system and of the apparatus and write
\be
\lb{D.1}
|u_j(t)\rangle = \hat a_j^\dagger (t) |0 \rangle \,\qquad 
|U_r(t)\rangle = \hat A_r^\dagger (t) |0 \rangle\,,
\ee
$\hat a_j^\dagger (t)$ and $\hat A_j^\dagger (t)$ being ordinary Heisenberg picture operators.

\quad Then let us assume the object system alone to be described at the initial time $t_a$ by
the statistical operator 
\bea
\lb{D.2}
&& \hat \rho ^{\rm O}(t_a)=  \sum_{ij} |u_i(t_a) \rangle 
 \rho _{ij}(t_a) \langle u_j(t_a)|= \nn \\
&& \qquad \qquad =\sum_{ij}   \hat a^\dagger (t_a)|0 \rangle  \rho _{ij}(t_a) \langle 0|
\hat a_i(t_a)\,.
\eea
The assumption the system to be small implies that the classic e. m. field stays negligible 
in the region occupied by the system until this does not come in contact with the apparatus. So at 
the time  $t_b$ we have
\bea
\lb{D.3}
 && \hat \rho ^{\rm O}(t_b)=\mathcal G(t_b,\,t_a) \{ \sum_{ij}\hat a_i^\dagger (t_a)|0 \rangle 
 \rho _{ij}(t_a) \langle 0|\hat a_j(t_a)\}= \nn \\
&& \qquad \qquad = \sum_{ij}   \hat a_i^\dagger (t_b)|0 \rangle  \rho _{ij}(t_b) 
\langle 0|\hat a_j(t_a)\,,  
\eea
where
\be
\lb{D.3a}
\rho_{ij}^{\rm O}(t_b)= \langle u_i(t_b)|\mathcal G(t_b,\,t_a) 
\{ \sum_{kl}\hat a_k^\dagger (t_a)|0 \rangle 
 \rho _{kl}(t_a) \langle 0|\hat a_l(t_a)\}|u_j(t_b \rangle \,.
\ee
\quad Similarly let be 
\be 
\lb{D.4}
\hat \rho^{\rm A}(t_a) = \sum_{rs} \hat A_r^\dagger (t_a)|0 \rangle  \rho _{rs}(t_a) 
\langle 0|\hat A_s(t_a)
\ee
the initial state of the apparatus. In this case we can assume that the counters remain in their 
charged states corresponding to the classical e. m. field having certain specific stable values
inside them until any interaction with some external object occurs. Again this corresponds to the 
classical world history of the electric field ${\bf E}_{\rm classic}(t,\,{\bf x})$ falling with 
certainty in a set $M_0 \in \Sigma_{t_a}^{t_b}$, being null the probability of occurrence of the 
complementary set  $M_0 '$.  Then 
\be
\lb{D.5}
\mathcal {F}(M_0;t_b \, t_a) \hat \rho ^{\rm A}(t_a) = \mathcal G(t_b,\, t_a) 
\hat \rho^{\rm A}(t_a)
\ee
and 
\be
\lb{D.6}
\hat \rho^{\rm A}(t_b) = \sum_{rs} \hat A_r^\dagger (t_b)|0 \rangle  \rho _{rs}^{\rm A}(t_b) 
\langle 0|\hat A_s(t_a)
\ee
with again
\be
\lb{D.7}
\rho_{rs}^{\rm A}(t_b)= \langle U_r(t_b)|\mathcal G(t_b,\,t_a) 
\left \{ \sum_{kl}\hat A_k^\dagger (t_a)|0 \rangle 
 \rho _{kl}^{\rm A}(t_a) \langle 0|\hat A_l(t_a)\right \}|U_s(t_b )\rangle \,.
\ee
\quad Since we have assumed that the object and the apparatus do not come in contact 
before  $t_b$, during the interval $(t_a,\, t_b)$ they must evolve independently and at the 
time $t_b$ for their compound state we have
\bea
\lb{D.8}
&& \hat \rho^{\rm T} (t_b)=\sum_{ij}\sum_{rs}\hat a_i^\dagger (t_b) \hat A_r^\dagger (t_b) |0 \rangle
\rho_{ij}^{\rm O}(t_b)\,\rho_{rs}^{\rm A}(t_b) \langle 0| \hat A_s (t_b) \hat a_j(t_b)= \nn \\
&& \qquad =\sum_{ij}\sum_{rs}|u_i,\,U_r;\,t_b \rangle
\rho_{ij}^{\rm O}(t_b)\,\rho_{rs}^{\rm A}(t_b) \langle u_j,\, U_s; t_b |\,.
\eea
\quad In a subsequent time interval $(t_b,\,t_c) $, as consequence of the interaction with the 
particles of the object system some of the counter shall discharge and every specific pattern 
of discharged counters is interpreted as corresponding to certain specific particles with specific 
energies and momenta present in the system. Then, if we denote by $N \in \Sigma_{t_b}^{t_c}$ the 
set of classical e. m. world histories corresponding to the parameters specifying the particles 
types, energies, momenta etc. falling in a certain set $T$, we have
\bea
\lb{D.9}
&& p(T,\,t_b)=P(t_c,\,t_b;\,N)=\nn \\
&&\qquad \qquad  \mathcal {\rm Tr} [\mathcal F(t_c,\,t_b;\,N)\,\hat \rho ^{\rm T}(t_b)]=
\sum _{jj'}F_{j'j}(T,\,t_b)\, \rho_{jj'}^{\rm O}(t_b)\,,
\eea
which is positive and can be confronted with (\ref{2.7}) and where obviously
\bea 
\lb{D.10}
&& F_{j'j}(T,\,t_b)= \sum_{ir} \langle u_i(t_b),\,U_r(t_b)|
\mathcal F(t_c,\,t_b;\,N) \left \{\sum_{ss'} |u_j(t_b),\,U_s(t_b)\rangle \right. \nn \\
&& \qquad \qquad \qquad \left.  \rho_{ss'}^{\rm A}(t_b) \langle u_j(t_b)',\,U_s'(t_b)|
\right \} |u_i(t_b),\,U_r(t_b)\rangle \,.
\eea
\quad To be more explicit let us assume that the vectors $|u_j(t_b)\rangle$ already correspond to 
a specifications of the of the state of the particles at the time $t_b$ and 
$N_j \in \Sigma_{t_b}^{t_c}$ the corresponding pattern of discharge of the counters, we can write 
\bea
&& \mathcal F(t_c,\,t_b;\,N_j)\,\hat \rho^{\rm T} (t_b) = \nn \\
&& \qquad = \rho_{jj}^{\rm O} (t_b) \, \mathcal G(t_c,\,t_b)
\{ \sum_{rs}|u_j,\,U_r;\,t_b \rangle \,\rho_{rs}^{\rm A}(t_b) \langle u_j,\, U_s; t_b |\rbrace\,.
\eea 
from which, since $\mathcal G(t_c,\,t_b) $ is trace-preserving, it follows  
\bea
&& p_j(t_b)\equiv P(t_c,\,t_b;\, N_j)= {\rm Tr} \{\mathcal F(t_c,\,t_b;\, N_j)\,
\hat \rho^{\rm T}(t_b)\} = \nn \\
&& \quad =\rho_{jj}^{\rm O}(t_b){\rm Tr} \{\sum_{rs}|u_j,\,U_r;\,t_b \rangle \,\rho_{rs}^{\rm A}(t_b) 
\langle u_j,\, U_s; t_b |\}=\rho_{jj}^{\rm O}(t_b)\,,
\eea
that is the prescription of usual elementary Quantum Theory up to the correction introduced in 
(\ref{D.4})
by the action of the mapping $\mathcal G(t_b,\,t_a)$.

%bbbbbbbbbbbbbbbbbbbbbbbbbbbbbbbbbbbbbbbbbbbbbbbbbbbbbbbbbbbbbb

\end{document}